\documentclass[a4paper, twocolumn]{scrartcl}
\usepackage[top=2.5cm, bottom=2.5cm, left=1.8cm, right=1.8cm]{geometry}
\usepackage{graphicx}
\usepackage{xcolor}
\usepackage{amsmath}
\usepackage{amssymb}
\usepackage{media9}
\usepackage[format=plain]{caption}
\usepackage{subcaption}
\usepackage{soul}
\usepackage[nolist]{acronym}
\usepackage{float}
\newcommand\textequal{%
\rule[.4ex]{4pt}{0.4pt}\llap{\rule[.7ex]{4pt}{0.4pt}}}
\newcommand{\testleft}{\leftarrow\!\shortmid}
\usepackage[colorlinks = true, linkcolor = blue,urlcolor  = blue, citecolor = blue, anchorcolor = blue]{hyperref}
\usepackage{authblk}

\usepackage{algorithm}
\usepackage{algpseudocode}

\PassOptionsToPackage{%
backend=bibtex8,bibencoding=ascii,%
language=auto,%
style=authoryear-comp, 
bibstyle=authoryear,dashed=false, 
sorting=nyt, 
maxbibnames=10, 
natbib=true 
,uniquename=init
,firstinits=true
,url=false
}{biblatex}
\usepackage{biblatex}

\title{Automatic Zig-Zag sampling in practice}

\author[1, 2, \S]{Alice Corbella}
\author[1]{Simon E. F. Spencer}
\author[1, 3]{Gareth O. Roberts}
\affil[1]{Department of Statistics, University of Warwick}\affil[2]{MRC Biostatics Unit, University of Cambridge}
\affil[3]{The Alan Turing Institute}
\affil[$\S$]{Corresponding author \href{mailto:alice.corbella@warwick.ac.uk}{\texttt{alice.corbella@warwick.ac.uk}}}
\date{\normalsize September 2022}
\begin{document}
\twocolumn[
\begin{@twocolumnfalse}
\maketitle
\begin{abstract}
Novel Monte Carlo methods to generate samples from a target distribution, such as a posterior from a Bayesian analysis, have rapidly expanded in the past decade. Algorithms based on Piecewise Deterministic Markov Processes (PDMPs), non-reversible continuous-time processes, are developing into their own research branch, thanks their important properties (e.g., super-efficiency). 

Nevertheless, practice has not caught up with the theory in this field, and the use of PDMPs to solve applied problems is not widespread. This might be due, firstly, to several implementational challenges that PDMP-based samplers present with and, secondly, to the lack of papers that showcase the methods and implementations in applied settings. Here, we address both these issues using one of the most promising PDMPs, the Zig-Zag sampler, as an archetypal example.

After an explanation of the key elements of the Zig-Zag sampler, its implementation challenges are exposed and addressed. Specifically, the formulation of an algorithm that draws samples from a target distribution of interest is provided. Notably, the only requirement of the algorithm is a closed-form {differentiable }function to evaluate the {log-}target density of interest, and, unlike previous implementations, no further information on the target is needed. 

The performance of the algorithm is evaluated against {canonical Hamiltonian Monte Carlo}, and it is proven to be competitive, in simulation and real-data settings. Lastly, we demonstrate that the super-efficiency property, i.e. the ability to draw one independent sample at a lesser cost than evaluating the likelihood of all the data, can be obtained in practice.

\vspace*{0.5cm}\textbf{Keywords}: automatic inference, Piecewise Deterministic Markov Processes, Zig-Zag sampler, gradient-based MCMC, super-efficiency

\end{abstract}
\vspace*{0.5cm}
\end{@twocolumnfalse}
]
\section{Introduction}\label{s.1}
Applications of Bayesian inference have proliferated immensely in the most disparate fields during the recent decades. The diffusion of Bayesian methods in several scientific communities owns its credit, among other things, to advances in software that allow one to draw samples from a posterior distribution $ p(\theta \vert y  ) $ of interest. The availability of programs such as \texttt{BUGS} \citep{gilks1994language} and \texttt{JAGS} \citep{plummer2003jags}, made standard \ac{MCMC} algorithms such as the Metropolis-within-Gibbs sampler available to the community and used in many applications.

In parallel to this proliferation of applications, the {methodology} behind \ac{MCMC} also expanded: recent research focussed on the exploitation of the gradient of the target density to explore the space in a more efficient manner. Early examples include the \ac{MALA} \citep{roberts1996exponential, roberts1998optimal}; and the \ac{HMC} algorithm \citep{neal2011mcmc}; these algorithms showed the practical gain in efficiency of exploiting information from the gradient. \ac{HMC} gained popularity in the 2010s thanks to the software \texttt{Stan} \citep{carpenter2017stan}, which has an embedded \ac{AD} tool that allows to draw samples from a target distribution, needing only the functional form of its \ac{pdf}.

More recently, algorithms based on \acp{PDMP} \citep{fearnhead2018piecewise} have been proposed and showed great potential \citep{bouchard2018bouncy, bierkens2018piecewise} thanks to their continuous-time behaviour and to convenient properties such as super-efficiency. Nevertheless, their use is not yet widespread, and very few papers use \ac{PDMP}-based algorithms to address Bayesian estimation problems \citep{koskela2022zig, chevallier2020reversible}. {Even fewer papers attempt to implement \ac{PDMP}-based algorithm in a general form \citep{pagani2020nuzz, bertazzi2021approximations}, unfortunately they don't retain exactness. }

This paper intends to help the practice to catch up with the advances in the theory in three ways: (i) it provides a lay explanation of the implementation of \ac{PDMP} algorithms, and specifically of the Zig-Zag sampler, making \acp{PDMP} available to a wider audience, both in terms of comprehension of the method and possibility of its  applications; (ii) it addresses some of the obstacles that prevent the use in practice of the algorithms for a general target density of interest, in particular, the availability of explicit form of the gradient of the target density and a bounding constant for it; and (iii) it provides examples of the use of \ac{PDMP} algorithms for real-data analyses.

Section \ref{s.2} introduces \acp{PDMP} in their general form and gives an example of a \ac{PDMP}-based algorithm: the canonical Zig-Zag sampler. This algorithm is used as a reference through the manuscript as its simple formulation makes illustration of many aspects of \acp{PDMP} clear and as it was shown to outperform other \ac{PDMP}-based algorithms in some applied settings  \citep{chevallier2020reversible}.
Section \ref{s.3} addresses the main implementation problems of the Zig-Zag sampler and provides the formulation of an algorithm that requires as input only a function that evaluates {a differentiable} target density at a specific point: the \textit{Automatic Zig-Zag} sampler. In Section \ref{s.4} the performance of this algorithm is evaluated against a competitive gradient-based scheme. Section \ref{s.5} provides examples of analyses carried out using Automatic Zig-Zag sampling. Section \ref{s.6} illustrates how super efficiency can be achieved in the context provided of the Automatic Zig-Zag sampler. Discussion and conclusions follow in Section \ref{s.7}.

\section{Background: the Zig-Zag sampler}\label{s.2}
A \ac{PDMP} is a continuous-time stochastic process denoted by $\boldsymbol{Z}_t $, which, in between random times, evolves according to deterministic dynamics. Values $ \boldsymbol{z}_t $ of the process can, for now, be thought of as $ d $-dimensional vectors with elements $ z^{(i)}_t $ for $i=1, \dots, d $. A \ac{PDMP} can be defined through specifying the following three components \citep{fearnhead2018piecewise}:
\begin{itemize}
	\item[(i)] a deterministic dynamic describing the change of the process over time which can be specified through an ordinary differential equation, 	
	\begin{equation}
		\label{eq2.1}\frac{\text{d} z^{(i)}_t}{\text{d}t} =\Phi_i(\boldsymbol{z}_t) \qquad \text{for }i=1, \dots, t.
	\end{equation}
	hence the state of the process at time $t+s $ can be computed as a deterministic function of the state of the process at time $ t $ and the elapsed time $ s $:
	$ z_{t+s}=\Psi (z_t, s)$; 
	\item[(ii)] random switching times which happen with rate dependent on the current state of the process $ \lambda(\boldsymbol{z}_t)$; and
	\item[(iii)] a transition kernel $ q(\cdot \vert \boldsymbol{z}_t) $ that determines the distribution of events that take place at the switching times and depends, again, on the current state of the process $ z_t $. 
\end{itemize}
The various  \ac{PDMP}-based algorithms differ among themselves in one or more of these specifics.

\subsection{Definition}
The Zig-Zag sampler is based on the simulation of a \ac{PDMP} composed of two, distinguishable, elements: a location $ \boldsymbol{X} \in \mathbb{R}^d $ and a velocity $ \boldsymbol{V} $. The velocity can be thought of as an auxiliary variable defined on the space $  \mathbb{V} = \{-1, +1\}^d $; the location instead, is typically the main component of interest: the sampler is constructed so that $\boldsymbol{ X }$ has stationary distribution with density $ \pi(\boldsymbol{x}) $ (e.g. a posterior density). Crucial to the definition of the Zig-Zag sampler is that the target density could be written as $ \pi(\boldsymbol{x})\propto e^{-U(\boldsymbol{x})} $, where $ U(\boldsymbol{x}) $ is sometimes called the \textit{potential}. 

Concerning the deterministic dynamics (i), the vector of velocities $ \boldsymbol{v} $ is assumed to be constant between switching times, with each dimension of $ \boldsymbol{x} $ increasing or decreasing at the same rate, so that Equation \eqref{eq2.1} is effectively:
\begin{equation}\label{eq2.2}
	\frac{\text{d}x_{t}^{(i)}}{\text{d}t}= v_t^{(i)}
\end{equation}
for $ i=1, \dots, d $. Given a starting state of the process $ (\boldsymbol{x_s}, \boldsymbol{v_s}) $, the velocity then switches according to (ii) the minimum of $ d $ \acp{NHPP} with rates
\begin{equation}\label{eq2.4}
	\lambda^{(i)}(t; \boldsymbol{x_s},\boldsymbol{v_s}) = \max\left\{{v_s}^{(i)} \frac{\partial}{\partial x^{(i)}
	}U(\boldsymbol{x_t}), 0\right\}.
\end{equation}
for $ i=1,2,\dots, d $, with $ \boldsymbol{x_t}= \boldsymbol{x_s}+\boldsymbol{v_s}\cdot t $ from (i). The intuition behind this formulation of the rate $ \lambda(\cdot) $ is similar to that of many other gradient-based scheme: if the {value} of the potential {is growing}, the chains is moving away from where the mass concentrates, and hence the direction changes. 

Lastly, (iii) the transition kernel $ q(\cdot\vert\boldsymbol{z_t}) $ is defined by the flipping operator $ F_m(\cdot) $ that inverts the sign of the $ m $-th dimension of the velocity, where $ m$ denotes the dimension of the earliest event in the realizations of the \acp{NHPP}.
\begin{equation}\label{eq2.5}
	F_m (v ^{(i)})=\begin{cases}
		-v^{(i)}\qquad &\text{ for }i=m\\
		v^{(i)} \qquad &\text{ for }i\neq m
	\end{cases}
\end{equation}
Bierkens et al 2019 \citep{bierkens2019zig} proved that a Zig-Zag process, under mild regularity conditions, converges to the invariant distribution of interest $ X $ with density $ \pi(x) $. 

To obtain the earliest realization of the $ d $ \acp{NHPP} with rates \eqref{eq2.4} it is possible sample from a one-dimensional inhomogeneous Poisson process with rate:
\begin{equation}\label{eq2.6}
	\lambda(t; \boldsymbol{x_s},\boldsymbol{v_s})=\sum_{i=1}^{d}	\lambda^{(i)}(t; \boldsymbol{x_s},\boldsymbol{v_s}) .
\end{equation}
The dimension in which the switch takes place is the realization of a Multinomial \ac{rv} with probabilities:
\begin{equation}\label{eq2.7}
	p_i	=\frac{\lambda^{(i)}(t; \boldsymbol{x_s},\boldsymbol{v_s}) }{\lambda(t; \boldsymbol{x_s},\boldsymbol{v_s})}
\end{equation}
for $ i=1, 2, \dots, d $.

\begin{figure}
	\centering
	\begin{subfigure}[b]{0.46\textwidth}
		\centering
		\vspace*{-0.3cm}	\includegraphics[scale=0.8]{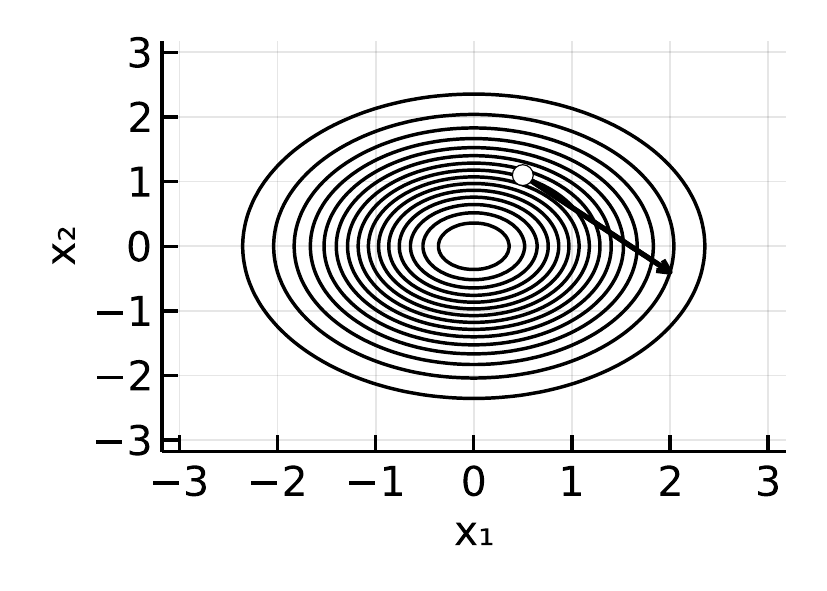}\vspace*{-0.5cm}
		\caption{\footnotesize$\boldsymbol{x_s}\textequal\begin{bmatrix}0.5\\1.1\end{bmatrix}, \boldsymbol{v_s}\textequal\begin{bmatrix}+1\\-1\end{bmatrix}$}
		\label{fig2.1a}
	\end{subfigure}
	\begin{subfigure}[b]{0.46\textwidth}
		\centering
		\vspace*{-0.3cm}	\includegraphics[scale=0.8]{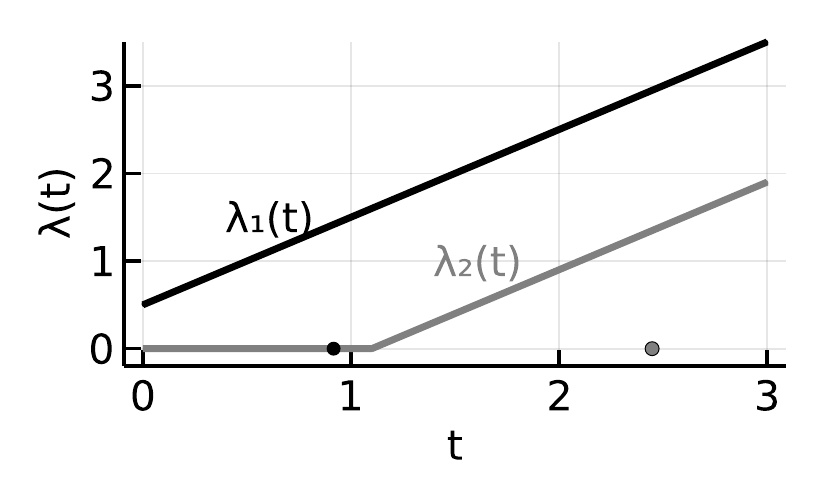}\vspace*{-0.5cm}
		\caption{\footnotesize$\lambda^{(1)}, \lambda^{(2)}$}
		\label{fig2.1b}
	\end{subfigure}
	\begin{subfigure}[b]{0.46\textwidth}
		\centering
		\vspace*{-0.3cm}	\includegraphics[scale=0.8]{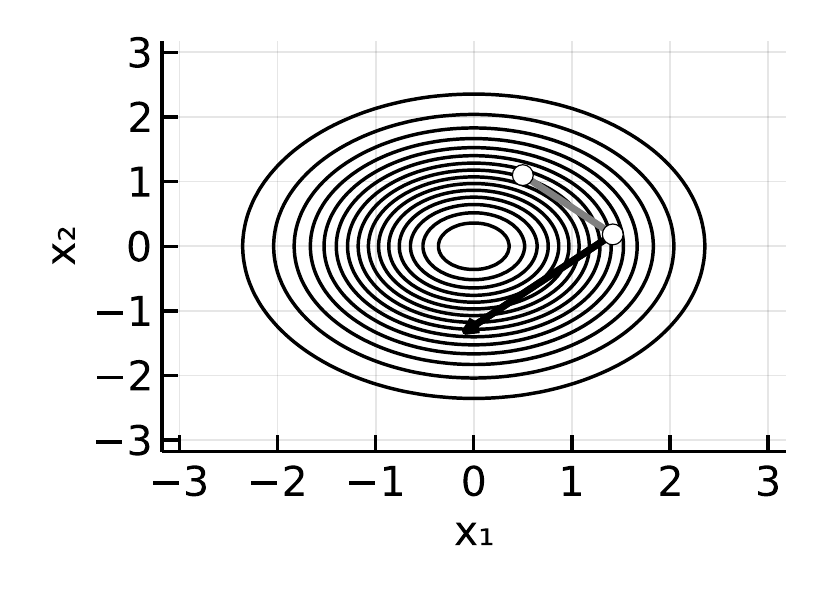}\vspace*{-0.5cm}
		\caption{\footnotesize$\boldsymbol{x_{t_1}}, \boldsymbol{v_{t_1}}\textequal\begin{bmatrix}-1\\-1\end{bmatrix}$}
		\label{fig2.1c}
	\end{subfigure}
	\begin{subfigure}[b]{0.46\textwidth}
		\centering
		\vspace*{-0.3cm}	\includegraphics[scale=0.8]{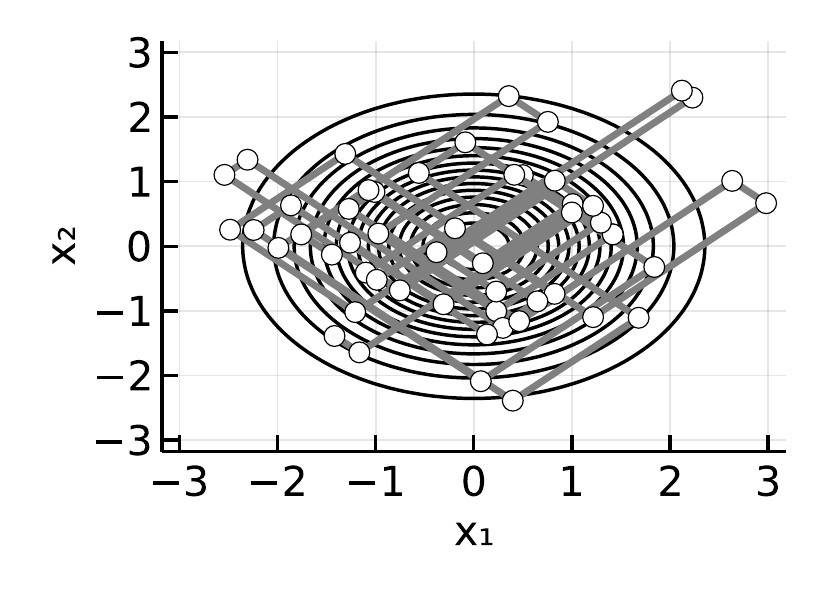}\vspace*{-0.5cm}
		\caption{\footnotesize $\boldsymbol{x_{t_0:t_L}}, \boldsymbol{v_{t_0:t_L}} {\color{white}\textequal\begin{bmatrix}-1\\-1\end{bmatrix}}$}
		\label{fig2.1d}
	\end{subfigure}
	\caption{Simulation of a Zig-Zag process targeting a bivariate independent standard normal distribution
		. (a) Initial location and velocity; (b) time-varying rate and samples from the \ac{NHPP} for dimension 1 (black) and 2 (grey); (c) first switching time; (d) first 50 switching times (white) and continuous-time sample (grey). }
	\label{fig2.1}
\end{figure}

An illustration of the first steps of the simulation of a Canonical Zig-Zag process is reported in Figure \ref{fig2.1}.

\subsection{Implementation}
\begin{figure}
	\centering
	\begin{subfigure}[b]{0.45\textwidth}
		\centering
		\hspace*{-0.7cm}\includegraphics{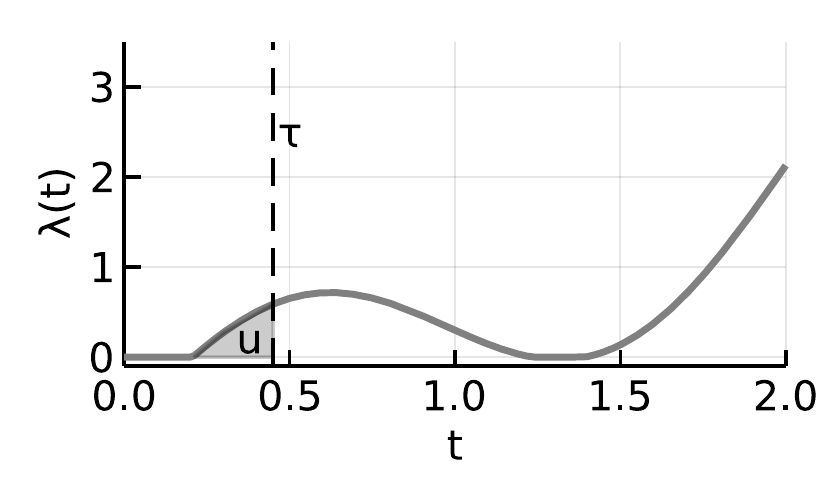}
		\caption{ $	\int_{0}^{\tau} \lambda(t) \text{d}{t}= u$\vspace*{0.7cm}}
		\label{fig2.2a}
	\end{subfigure}
	\begin{subfigure}[b]{0.45\textwidth}
		\centering
		\hspace*{-0.7cm}\includegraphics{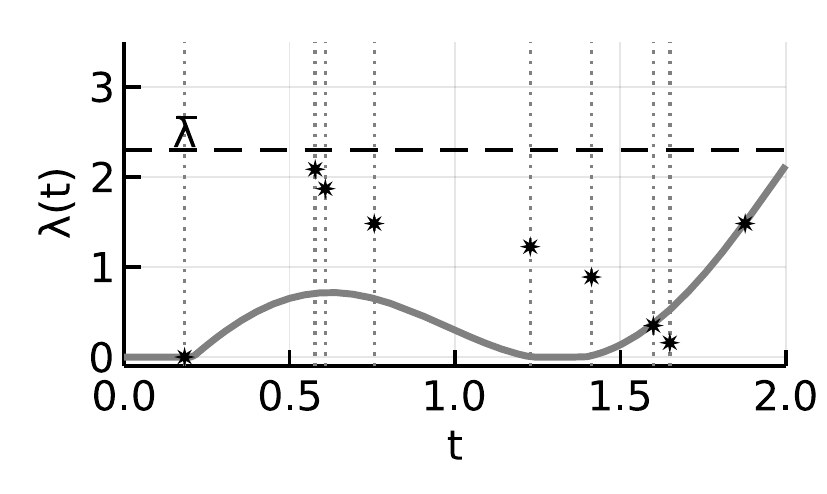}
		\caption{ Samples from the \ac{HPP} (dotted vertical lines) with rate $ \overline{\lambda} $ (dashed) accepted with probabilities $ \frac{\lambda(\tau^*)}{\overline{\lambda}} $ (stars).   }
		\label{fig2.2b}
	\end{subfigure}
	\caption{Sampling mechanisms for a \ac{NHPP}: (a) time-scale transformation and (b) thinning method. }
	\label{fig2.2}
\end{figure}

The practical implementation of the algorithm {requires} sampling from an \ac{NHPP} with rate $ \lambda(t)$, where arguments $\boldsymbol{x_s}$ and $\boldsymbol{v_s} $ are omitted since they are constant between switching times. As summarised by \cite{lewis1979simulation}, this can be done either via time-scale transformation, finding $ \tau $ such that:
\begin{equation}\label{eq2.8}
	\int_{0}^{\tau} \lambda(t) \: \text{d}{t}= u
\end{equation}
given $ u$ sampled from an $ \text{Exp}(1) $; or via thinning, i.e. (i) finding a constant upper bound $ \overline{\lambda} $ such that $ \overline{\lambda}\geq\lambda(t)$, either globally $ \forall t $ or in some interval $ [a,b]$, (ii) sampling a candidate point $ \tau^* $ from an \ac{HPP} with rate $ \overline{\lambda} $ and (iii) accepting the candidate point with probability $ \frac{\lambda(\tau^*)}{\overline{\lambda}} $. These sampling techniques are illustrated in Figure \ref{fig2.2}.

Analytically determining the point $ \tau $ that satisfies Equation \eqref{eq2.8} is often impossible, above all due to the maximum contained in \eqref{eq2.4}. Solving Equation \eqref{eq2.8} numerically is often more expensive than finding a suitable upper bound $ \overline{\lambda} $ and simulating the process via thinning: while the latter requires only a limited, wisely chosen, number of evaluations of the objective function, numerical integration implies a discretization of the domain $ t $ and the evaluation of the function at these numerous discrete points. Hence, here the thinning method is used to simulate an \ac{NHPP}. 

Using the ingredients of Equations \eqref{eq2.2}, \eqref{eq2.4}, and \eqref{eq2.5}, it is possible to obtain the positions and the velocity of the process at each switching time $ t_k $: $ \left\lbrace \boldsymbol{x_{t_k}},\boldsymbol{v_{t_k}} \right\rbrace _{k=1}^K$. These are called the \textit{skeleton points} of the sampled distribution. The value of the process at each time $ t $ between two skeleton points can then be obtained using the deterministic dynamics of Equation \eqref{eq2.2} which results in:
\begin{equation*}
	\boldsymbol{x_t}=\boldsymbol{x_{t_k}}+\boldsymbol{v_{t_k}}(t-t_k) \qquad \text{ for } t \in [t_{k}, t_{k+1}].
\end{equation*}

The pseudo-code of the Zig-Zag sampler with thinning when a global upper bound $ \overline{\lambda} $ is known, is reported in Algorithm \ref{alg1}.

\begin{algorithm}
	\caption[Canonical Zig-Zag algorithm]{Canonical Zig-Zag algorithm with thinning with known global bound $ \overline{\lambda} $. }\label{alg1}
	\begin{algorithmic}[1]
		\Require Initial velocity $ \boldsymbol{v_0} $; initial location $ \boldsymbol{x_0} $; total number of skeleton points to be sampled $ K $; functions to compute the dimension-wise and global switching rate $ \lambda(\cdot) $ and $\lambda^{(i)}(\cdot)$; an upper bound for the global rate $ \overline{\lambda} $ .
		\Ensure Time, location and velocity at $ K $ skeleton points: $ \left\lbrace t_k,  \boldsymbol{x_k} , \boldsymbol{v_k}\right\rbrace_{k=1}^K $ 
		\State $ t_0=0  $ \hspace*{\fill}\#set starting time
		\State $\textsc{t}\testleft t_0,\textsc{ x}\testleft\boldsymbol{x_0},  \textsc{v}\testleft\boldsymbol{v_0}$
		
		\hspace*{\fill}\#set current state of the process 
		\While{$ k\leq K $}
		\State $ \tau^* \sim \text{Exp}(\overline{\lambda}) $\hspace*{\fill}\#propose switching time
		\State $  \lambda(\tau^*)=\sum_{i=1}^{d}	\lambda^{(i)}(\tau^*; \textsc{x},\textsc{v}) $
		
		\hspace*{\fill}\# compute the global rate at $  \tau^* $
		\State $ u \sim \text{Ber} ( \frac{\lambda(\tau^*)}{\overline{\lambda}} )$\hspace*{\fill}\# accept/reject $  \tau^* $
		
		\If{$ u=1 $}
		\State $ \textsc{t}\testleft\textsc{t}+\tau^* $ \hspace*{\fill}\#progress time
		\State 
		$ \textsc{x}\testleft\textsc{x}+\textsc{v}\tau^* $\hspace*{\fill}\#progress location
		\State $  m\sim \text{Multinom}\left(1:d ; \left\lbrace\frac{ \lambda^{(i)}(\tau^*; \textsc{x},\textsc{v})}{\lambda(\tau^*)} \right\rbrace_{i=1}^d \right)$
		
		\hspace*{\fill}\#sample component
		\State $ \textsc{v}\testleft F_m(\textsc{v})$\hspace*{\fill}\#flip velocity of dim $ m $
		\State $ t_k\testleft\textsc{t}, \boldsymbol{v_k}\testleft\textsc{v}, \boldsymbol{x_k}\testleft\textsc{x} $
		
		\hspace*{\fill}\#save skeleton point
		\State $ k=k+1 $
		\Else
		\State $ \textsc{t}\testleft\textsc{t}+\tau^* $ \hspace*{\fill}\#progress time
		\State$ \textsc{x}\testleft\textsc{x}+\textsc{v}\tau^* $\hspace*{\fill}\#progress location
		\State$ \textsc{v}\testleft\textsc{v}$\hspace*{\fill}\#retain velocity
		\EndIf
		\EndWhile
	\end{algorithmic}
\end{algorithm}

\subsection{Beyond Canonical Zig-Zag sampling}
The Canonical Zig-Zag algorithm is not the only example of the use of \acp{PDMP} to sample from a target density of interest $ \pi(\boldsymbol{x}) $. The basic algorithm can be changed and extended in a number of ways to improve its performance on specific targets; moreover, different deterministic dynamics and switching rates/kernels can been used to formulate other PDMP-based algorithms (see, for example \cite{bouchard2018bouncy}, \cite{wu2020coordinate}, and  \cite{bierkens2020boomerang}). Nevertheless, the focus of this paper is on the Canonical Zig-Zag algorithm to provide a simple example where our methods are applicable. 

\subsubsection{Non-canonical Zig-Zag sampling algorithms}
The switching rate in Equation \eqref{eq2.4} could be further extended by  adding an excess switching rate $ \gamma^{(i)}(\boldsymbol{x_t}, \boldsymbol{v_t}) $ such that \begin{equation}\label{eq2.9}
	\begin{cases}
		\gamma^{(i)}(\boldsymbol{x_t}, \boldsymbol{v_t})&\geq 0 \\
		\gamma^{(i)}(\boldsymbol{x_t}, \boldsymbol{v_t})&= \gamma^{(i)}(\boldsymbol{x_t}, F_i(\boldsymbol{v_t})) .
	\end{cases}
\end{equation}
leading to switching rate:
\begin{equation}\label{eq2.10}
	\begin{split}
		\lambda^{(i)}(t; \boldsymbol{x_s},\boldsymbol{v_s}) = &\max\left\{{v_s}^{(i)} \frac{\partial}{\partial x^{(i)}
		}U(\boldsymbol{x_t}), 0\right\}\\
		&\qquad+\gamma^{(i)}(\boldsymbol{x_t}, \boldsymbol{v_t}) 
	\end{split}
\end{equation}
for $ i=1, \dots, d $. 

This simple modification, discussed in \cite{bierkens2019zig}, allows the process to still converge to the correct target distribution \citep{bierkens2019ergodicity} and slightly increases the event rate, generating extra switching times in addition to those driven by the potential $ U(\boldsymbol{ x }) $. These switches are often called \textit{refreshments} and, while in principle adding {excessive }refreshments will impoverish the mixing of the process \citep{andrieu2021peskun}, many interesting constructs such as the Zig-Zag with subsampling, can be built by considering refreshment switches. 

Other extensions have been formulated, in order to improve the performance of the Zig-Zag sampler on specific distributions/applications (e.g. heavy tailed distributions, highly correlated distributions, variable selection problems, etc.).  One of these extension proposed the addition of moves beyond the flipping operator or the extension of the velocity domain beyond $ \mathbb{V}=\left\lbrace-1;+1 \right\rbrace^d  $ (see for example \cite{chevallier2020reversible}). \cite{vasdekis2021speed} proposed the use of a function $ S(\boldsymbol{x_t}) $ that allows the acceleration of the process according to its position (e.g. speeding up in the tails). In a recent work \citep{bertazzi2020adaptive}, an adaptive version of the Zig-Zag sampler and other \ac{PDMP} algorithms was proposed, whereby the velocity is changed so that the performance of the algorithm would be equal to that of the canonical Zig-Zag sampler on an isotropic Gaussian distribution. This was proven to substantially improve efficiency.

%
%
%
%
%
\section{Automatic Zig-Zag sampling}\label{s.3}
This section describes some methods to allow the \textit{automatic} use of the Zig-Zag process. Here \textit{automatic} means that the only input needed is a {differentiable }functional form for the potential $ U(\boldsymbol{x})= -\log(\pi(\boldsymbol{x})) +c$, where $ \pi(\boldsymbol{x}) $ is the target density. Note that this goal, not only implies that manual differentiation of $ U(\boldsymbol{x}) $ should not be needed prior to start the analysis, but also that the algorithm should be run (i.e. produce a sample from the \ac{PDMP}) without relying on any external information about properties of the density such as its concavity or bounds.

\subsection{Automatic Differentiation}\label{s.3.1}
AD is a set of techniques that, given a function $ f(\boldsymbol{x}):\mathbb{R}^n \to \mathbb{R}^m $, allows the evaluation of $ f'^{(i)} ({\boldsymbol{x_0}})$, the derivative of $ f $ for a specific point $ \boldsymbol{x_0} \in \mathbb{R}^n $ w.r.t dimension $ i=1,\dots, n $ \citep{baydin2018automatic}. 
Notably, Automatic differentiation, not only provides an exact solution, but also it tends to be efficient: following the \textit{Cheap Gradient Principle}, the computational cost of computing the gradient of a scalar-valued function is nearly the same (often within a factor of 5) as that of simply computing the function itself \citep{griewank2008evaluating}.

The basis of Automatic Zig-Zag sampling is in computing the rate at Equation \eqref{eq2.4} via \ac{AD} for the point  $ \boldsymbol{x_t}= \boldsymbol{x_s}+\boldsymbol{v_s}\cdot t $ whenever needed; 
Algorithm \ref{alg1} follows identically as before. 

Since \ac{AD} does not introduce any numerical approximation, all results proven for the Zig-Zag sampler (e.g. the main convergence statements of \cite{bierkens2019zig}) hold for the Automatic Zig-Zag sampler.

\subsection{Rate bounds}\label{s.3.2}
In the practical implementation of the Automatic Zig-Zag sampler, the main challenge is to find an upper bound for the global rate $ 	{\lambda}(t) $  of the \ac{NHPP}. While a global or local upper bound to the gradient of $ U(\boldsymbol{ x }) $ might be known for many distributions of interest, we are looking for a general method that could bound, at least locally, any closed-form density of interest.

Constant upper bounds are used here and should be found under the consideration that if the upper bound is too large, then a large amount of computational effort is wasted in sampling candidate skeleton points (and evaluating $ {\lambda^{(i)}}(t) $) that are then rejected. Therefore, the upper bound should be as close as possible to the time-varying rate $ 	{\lambda}(t) $. Hence, a pragmatic  approach is chosen: the rate bound is defined locally (i.e. specific to the current location and velocity of the process) to be the maximum of the global rate in an interval of size $ t_\textsc{max} $:
\begin{equation}\label{eq3.2}
	\overline{\lambda}(t_\textsc{max}, \boldsymbol{x_s},\boldsymbol{v_s})=\max_{t \in [0,t_\textsc{max} ]} 	\left\lbrace {\lambda}(t; \boldsymbol{x_s},\boldsymbol{v_s}) \right\rbrace 
\end{equation} 
which, for brevity is denoted by $ \overline{\lambda} $, dropping the notation of the local dependence. 
If no events are sampled in the \ac{NHPP} in the interval $ [0,t_\textsc{max}] $, then the Zig-Zag process jumps straight to $ \boldsymbol{z_{s+t_\textsc{max}}}=( \boldsymbol{x_s}+\boldsymbol{v_s}\cdot t_\textsc{max} , \boldsymbol{v_s}$) without any further evaluations of the rates. The rate bound is then re-evaluated for the next interval and sampling continues. Values of $ t_\textsc{max}  $ are further discussed in Section \ref{s.3.3}.

Since ${\lambda}(t) $ consists of a blackbox and there is no explicit form of the rate function, finding an analytical maximum is impossible. 
Among the numerical optimization methods, gradient- and Hessian-free methods are particularly attractive since they are highly efficient and robust for univariate optimization problems, such as this one.

\subsubsection{Brent's {optimization} method}\label{s.3.2.1}
Similarly to other univariate optimization methods, the goal of this routine is to obtain the minimum of an objective function $ f: \mathbb{R}^1 \to  \mathbb{R}^1 $ (if the maximum is needed, as in this case, the optimization routine is run on $ -f $ instead).
Brent's method \citep{vetterling1992numerical} combines inverse parabolic interpolation with Golden Section search \citep{kiefer1953sequential}. 

Parabolic interpolation starts from three points $ (a, f(a)), (b, f(b)), (c, f(c)) $ such that $ a<b<c $, $ f(b)\leq f(a)$ and $f(b)\leq f(c) $, and finds the abscissa $ x $ of the vertex of a parabola interpolating the three points via the formula:
\begin{equation}\label{eq3.3}
	x=b-\frac{1}{2}\frac{(b-a)^2[f(b)-f(c)]-(b-c)^2[f(b)-f(a)]}{(b-a)[f(b)-f(c)]-(b-c)[f(b)-f(a)]}
\end{equation}
Substituting the highest point among $ (a, f(a)), (b, f(b)), (c, f(c)) $ with $ (x,f(x)) $ and iterating this formula, until a fixed tolerance is reached, should approach the minimum of the function $ f $. 

The  Golden Section search brackets the minimum of $ f(x) $ with intervals that are chosen to respect the golden ratio $\frac{1+\sqrt{5}}{2} $, so that their width can be reduced most efficiently. 

The Brent method combines these two methods by keeping track of 6 points:
\begin{itemize}\setlength{\itemsep}{-2pt}
	\item[$ a $ / $ b $] lowest/highest point of the interval bracketing the minimum
	\item[$ x $] best candidate minimum point found so far
	\item[$ v $] point with the second least value found so far
	\item[$ w $] value of $ v $ at the previous iteration
	\item[$ u $] point of the most recent evaluation of $ f $
\end{itemize}
The optimization scheme is as follows:
\begin{enumerate}\setlength{\itemsep}{-2pt}
	\item Propose a new point $ x^* $ by parabolic interpolation with Equation \eqref{eq3.3} on $ (x, f(x)), (v, f(v))$ and $ (w, f(w)) $
	\item \begin{itemize}\setlength{\itemsep}{-2pt}
		\item[\textbf{if}] the new point lies in the bracketing interval: $ a\leq x^*\leq b$
		\item[\textbf{and}] convergence is obtained by steps that are increasingly smaller $ \vert f(x)-f(x^*)\vert\leq 0.5 \vert f(v)-f(w)\vert $
	\end{itemize}
	accept the new proposed point and uprate the bracketing interval to either $ (a, x) $ or $ (x, b) $
	\item otherwise update the bracketing interval by Golden Search.
\end{enumerate}
These steps are iterated until some tolerance is reached. 

Note that the Golden Search method is slow and highly reliable, while polynomial interpolation is much quicker but is founded on the assumption that the function has an approximately parabolic behaviour. Brent's method would then be at worst as slow as Golden Search method.

\subsubsection{Modification for Zig-Zag}\label{s.3.2.2}
In the application considered here, Brent's optimization method is used to solve Equation  \eqref{eq3.1} and obtain a maximum. In this context, a few considerations can be made:
\begin{itemize}\setlength{\itemsep}{-2pt}
	\item[(i)] If the distribution considered is unimodal, the rates  \eqref{eq3.1} will be often monotonic;
	\item[(ii)] If $ t_\textsc{max}  $ is chosen to be smaller than the distance to the nearest mode, even in the case of a multimodal distribution, the rates would be mostly monotonic in the optimization interval $ [0,t_\textsc{max} ] $;
	\item[(iii)] If the function to be maximised is monotonic in the interval $ [0,t_\textsc{max} ] $, the maximum is either at $ 0 $ or at $ t_\textsc{max}  $.
\end{itemize}
Given these considerations, Brent's method can be modified and computations can be shortened after some tests for monotonicity. For this reason, a modification to Brent's method is proposed: after the first iteration is carried out, a check is run to assess if any of the two limits of the bracketing interval are unchanged. If so, then a second check is performed to confirm that the rate function approaches the end of the interval from below, by evaluating $ {\lambda}(t; \boldsymbol{x_s},\boldsymbol{v_s}) $ a distance $ \varepsilon $ from the end, for some small $ \varepsilon> 0$.  If this is the case, the rate is assumed to be monotonic in $ [0,t_\textsc{max} ] $ and the value of the rate at the selected limit is taken as upper bound $ \overline{\lambda}$; alternatively Brent's algorithm is run until convergence to the resulting maximum $ x $ and set $ \overline{\lambda}= x $. 

\subsection{Tuning of $ t_{\text{\tiny MAX}} $}\label{s.3.3}
With Equation \eqref{eq3.3}, a parameter $ t_\textsc{max} $ is introduced into the Automatic Zig-Zag algorithm. This is effectively a tuning parameter, with $ \overline{\lambda} $ being more or less \textit{local} according to the magnitude of $ t_\textsc{max}  $. 

\begin{figure}[!h]
	\centering 
	\begin{subfigure}[b]{0.45\textwidth}
		\centering
		\hspace*{-0.7cm}\includegraphics{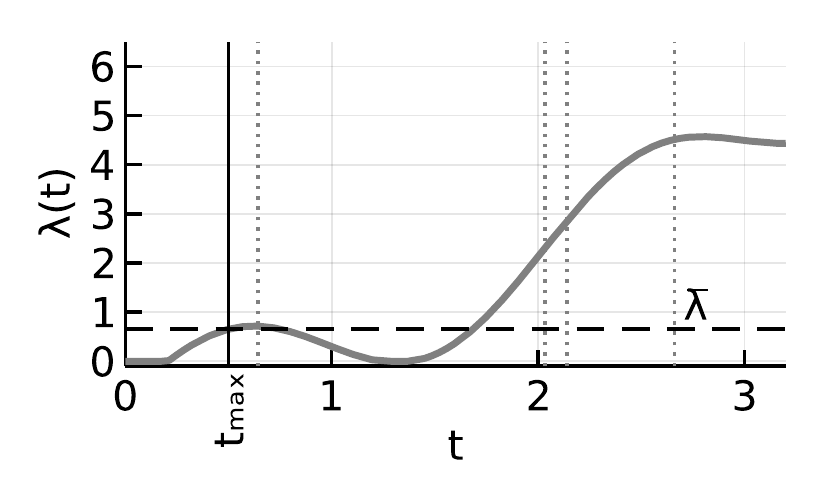}
		\caption{ Small $ t_\textsc{max} $}
		\label{fig3.1a}
	\end{subfigure}\hfill
	\begin{subfigure}[b]{0.45\textwidth}
		\centering
		\hspace*{-0.7cm}\includegraphics{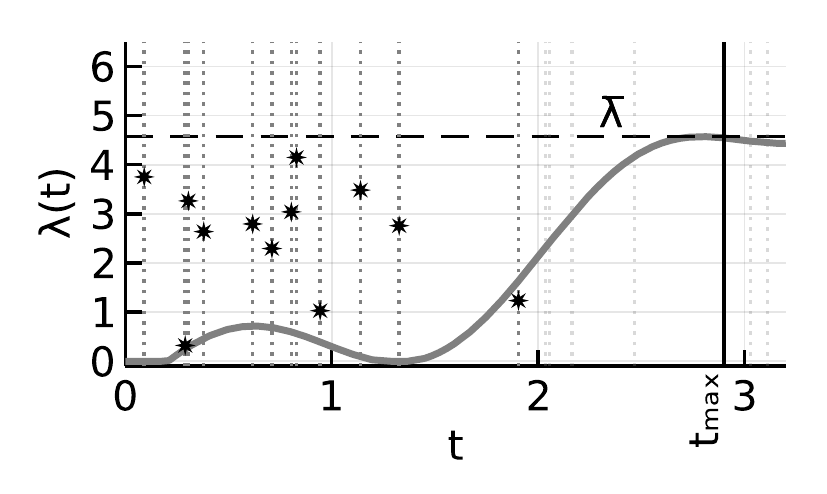}
		\caption{Large $ t_\textsc{max} $}
		\label{fig3.1b}
	\end{subfigure}
	\caption{Tuning of $ t_\textsc{max} $. (a) A deterministic step encouraged by the choice of a small value. (b) Many proposals of \ac{HPP} prior to acceptance, caused by a very large value of the local bound.  
	}
	\label{fig3.1}
\end{figure}

When $ t_\textsc{max}  $ is small, $ \overline{\lambda} $ would be very local, with ${\lambda}(t; \boldsymbol{x_s},\boldsymbol{v_s}) $ varying little in the interval, the rate should be smaller, hence the \ac{HPP} proposal events should be more rare, making it more likely for the \ac{PDMP} to reach $ t_\textsc{max} $ without any switch; every time this happens, another optimization step needs to be run to obtain a new bound $ \overline{\lambda} $. On the other hand, if $ t_\textsc{max} $ is very large, \ac{HPP} events are likely to be proposed more often, and for all the proposed times the rate $ {\lambda}(s)  $ has to be evaluated. An illustration of this tuning criterion can be found in Figure \ref{fig3.1}. 

The optimal $ t_\textsc{max}  $ is chosen by minimizing the number of evaluations of the rate $ {\lambda}(s) $ per switching time, which includes both the evaluations within the optimization algorithm and the computation of the acceptance probabilities. This can be done via some preliminary runs of the algorithm.  

The pseudo-code of the Automatic Zig-Zag sampling taking as input a value of $ t_\textsc{max} $ is reported in Algorithm \ref{alg2}.

\begin{algorithm*}
	\caption[Automatic Zig-Zag algorithm]{Automatic Zig-Zag algorithm with thinning and local optimization.}\label{alg2}
	\begin{algorithmic}[1]
		\Require Initial velocity $ \boldsymbol{v_0} $; initial location $ \boldsymbol{x_0} $; total number of skeleton points to be sampled $ K $; functions to compute the dimension-wise and global switching rate $ \lambda(\cdot) $ and $\lambda^{(i)}(\cdot)$; $t_\textsc{max} $ the tuning parameter.
		\Ensure Time, location and velocity at $ K $ skeleton points: $ \left\lbrace t_k,  \boldsymbol{x_k} , \boldsymbol{v_k}\right\rbrace_{k=1}^K $ 
		\State $ t_0=0 , k=1 $ \hspace*{\fill}\#set starting time and skeleton count
		\State $\textsc{t}\testleft t_0,\textsc{ x}\testleft\boldsymbol{x_0},  \textsc{v}\testleft\boldsymbol{v_0}$
		\hspace*{\fill}\#set current state of the process 
		\State $ \overline{\lambda}=\max_{t \in [0,t_\textsc{max} ]} 	\left\lbrace {\lambda}(t; \textsc{x}, \textsc{v}) \right\rbrace $	\hspace*{\fill}\#compute upper bound 		
		\State $ \tau^* \sim \text{Exp}(\overline{\lambda}) $\hspace*{\fill}\#propose switching time
		\State $ \tau^{opt}\testleft \tau^*  $
		\hspace*{\fill}\#track time from last optimization
		\While{$ k\leq K $}
		\State $ u=0 $\hspace*{\fill}\#set acceptance to 0 until next proposal
		\While{$ \tau^{opt}\leq t_\textsc{max}$ \textbf{and} $ u=0 $}
		\State $  \lambda(\tau^{opt})=\sum_{i=1}^{d}	\lambda^{(i)}(\tau^{opt}; \textsc{x},\textsc{v}) $
		\hspace*{\fill}\# compute the global rate at $  \tau^{opt} $
		\State $ u \sim \text{Ber} ( \frac{\lambda(\tau^{opt})}{\overline{\lambda}} )$\hspace*{\fill}\# accept/reject $  \tau^{opt} $
		\If{$ u=1 $}
		\State $  m\sim \text{Multinom}\left(1:d ; \left\lbrace\frac{ \lambda^{(i)}(\tau^{opt}; \textsc{x},\textsc{v})}{\lambda(\tau^*)} \right\rbrace_{i=1}^d \right)$
		\hspace*{\fill}\#sample component to switch
		\State $ \textsc{t}\testleft\textsc{t}+\tau^{opt} $ \hspace*{\fill}\#progress time
		\State 
		$ \textsc{x}\testleft\textsc{x}+\textsc{v}\tau^{opt} $\hspace*{\fill}\#progress location
		\State $ \textsc{v}\testleft F_m(\textsc{v})$\hspace*{\fill}\#flip velocity of dimension $ m $
		\State $ t_k\testleft\textsc{t}, \boldsymbol{v_k}\testleft\textsc{v}, \boldsymbol{x_k}\testleft\textsc{x} $		
		\hspace*{\fill}\#save skeleton point
		\State $ k =k+1 $\hspace*{\fill}\#increase skeleton count
		\State $ \overline{\lambda}=\max_{t \in [0,t_\textsc{max} ]} 	\left\lbrace {\lambda}(t; \textsc{x}, \textsc{v}) \right\rbrace $	\hspace*{\fill}\#compute new upper bound from switch
		\State $ \tau^* \sim \text{Exp}(\overline{\lambda}) $\hspace*{\fill}\#propose switching time
		\State $ \tau^{opt}\testleft\tau^*  $
		\hspace*{\fill}\#reset time from last optimization
		\Else
		\State $ \tau^* \sim \text{Exp}(\overline{\lambda}) $\hspace*{\fill}\#sample new time increment
		\State $ \tau^{opt}\testleft  \tau^{opt}+\tau^*  $\hspace*{\fill}\#compute new switching proposal
		\EndIf
		\EndWhile
		\If{$ \tau^{opt}> t_\textsc{max}$ \textbf{and} $ u=0 $}\hspace*{\fill}\#if the horizon is reached with no switch
		\State $ \textsc{t}\testleft\textsc{t}+t_\textsc{max}$ \hspace*{\fill}\#progress time deterministically until horizon
		\State 
		$ \textsc{x}\testleft\textsc{x}+\textsc{v}t_\textsc{max} $\hspace*{\fill}\#progress location deterministically until horizon
		\State $ \textsc{v}\testleft \textsc{v}$\hspace*{\fill}\#retain velocity
		\State $ \overline{\lambda}=\max_{t \in [0,t_\textsc{max} ]} 	\left\lbrace {\lambda}(t; \textsc{x}, \textsc{v}) \right\rbrace $	\hspace*{\fill}\#compute new upper bound from new location
		\State $ \tau^* \sim \text{Exp}(\overline{\lambda}) $\hspace*{\fill}\#propose switching time
		\State $ \tau^{opt}\testleft\tau^*  $
		\hspace*{\fill}\#reset time from last optimization
		\EndIf
		\EndWhile
	\end{algorithmic}
\end{algorithm*}

\section{Performance evaluation}\label{s.4}
This section investigates the performance of the Automatic Zig-Zag sampler.
The performance is tested on some bivariate distributions starting from an uncorrelated bivariate normal and exploring increasingly-more-challenging features. 
Main results are reported in Section \ref{s.4.3} and an exhaustive description of each simulation is reported in the Supplementary Information. 

\subsection{Performance metrics}\label{s.4.2}
Performance is evaluated according to two criteria: efficiency and robustness. 
\subsubsection{Efficiency}
To measure efficiency, the \ac{ESS} of the sample drawn with the two algorithms is compared; the samplers are run given a specific budget. 
The computational budget $ c$ is defined as the total number of evaluations of the gradient of the minus-log density of the target distribution ($ \nabla U(\boldsymbol{x}) $). 

For the Automatic Zig-Zag algorithm, the number of gradient evaluations required to produce each skeleton point comprises, for skeleton point $ k $: $ C^{\textsc{opt}}_k $, the number of evaluations of the switching rate during the optimization routine to find the bound $ \overline{\lambda} $; and $ C^{\textsc{tpp}}_k$, the number of proposed times for the thinned Poisson process. The number of evaluations over all the sampled skeleton is:
\begin{equation}\label{key}
	C^{\textsc{zz}} =\sum_{k=1}^{K}  \left\lbrace C^{\textsc{opt}}_k +C^{\textsc{tpp}}_k\right\rbrace 
\end{equation}
and therefore, the sampler stops at the smallest $ K$ such that $ C^\textsc{zz}\geq c $.

For a canonical \ac{HMC} algorithm that performs $ L $ leapfrog steps per iteration and $ K $ iterations, the number of evaluation of the gradient is: 
\begin{equation}\label{eq4.2}
	C^{\textsc{hmc}} =(L+1)\times K .
\end{equation}
Hence the sampler is run for $K=\frac{c}{L+1} $ steps.

The Automatic-Zig-Zag efficiency is computed using the \ac{ESS} for continuous-time trajectories presented in \cite{bierkens2019zig} (Supplementary Information S.2) for  the function $ h(\boldsymbol{x})=x_i $ for all the $ i $ coordinates. Similarly, the batch-means approach for \ac{ESS} calculated from  discrete-time samples is used to evaluate the efficiency of the runs of the \ac{HMC} algorithm. To summarise the results in \ac{ESS} across multiple dimensions, it is useful to compare the dimension with smallest \ac{ESS} (Median \ac{ESS} over 100 independent chains) since this dimension mixes more slowly and hence constrain the chain to an overall slower mixing.

\subsubsection{Robustness}
The other aspect examined to assess the performance of the Automatic Zig-Zag sampler was whether or not the algorithm was robust with respect to particular features of the distribution (e.g. heavy or light tails, multimodality). 

In particular, the ability of a tuned algorithm to properly explore the target distribution was investigated, even when starting from location far away from the mode. This was conducted mainly graphically and robustness was assessed qualitatively.

\subsection{Simulation set up} \label{s.4.1}
The Automatic Zig-Zag algorithm is compared with the Canonical \ac{HMC} algorithm (for a description of the latter see Section 3 of \cite{neal2011mcmc} or Section S1 of the Supplementary Information). The \ac{HMC} algorithm is said to be \textit{canonical} when, in the velocity-position framework similar to the one defined above, the velocity is sampled from an independent multivariate Normal distribution. This is a rigid structure, compared to other versions of the \ac{HMC} algorithm that choose a velocity distribution optimally with respect to the target density. 
Similarly, the version of Zig-Zag sampler used here is the canonical Zig-Zag, which employs constant velocities in $ \left\lbrace -1, +1 \right\rbrace^d  $, with no attempt to choose an optimal velocity that matches to the target distribution. 

Both algorithms are tuned before the comparison via preliminary runs. More specifically, $ t_\textsc{max} $ is chosen according to the criterion explored in Section \ref{s.3.3}, while the choice of the tuning parameters of the \ac{HMC} (i.e. the total integration time $L\times\varepsilon $ and of the number of leapfrog steps $ L $) is known to be a troublesome task \citep{sherlock2021apogee}. The procedure adopted here for tuning includes many graphical assessments and is reported in Section S1 of the Supplementary Information. 
\subsection{Results}\label{s.4.3}
The results of the efficiency analysis on various forms of Bivariate Gaussian distribution are reported in Table \ref{t1}. 
The algorithms were tested on an isotropic Gaussian distribution (\texttt{IsoG2}); on a bivariate Gaussian distribution where the two components had the same scale and high correlation $ \rho=0.9 $ (\texttt{CorG2}); on a bivariate Gaussian distribution with independent components with very different scales $ \sigma^2_1=1, \sigma^2_2=100 $ (\texttt{DscG2}); and on a bimodal distribution, a mixture of Gaussians (\texttt{BimodG2}). 

\begin{table}[h]
	\centering
	\caption{Smallest \ac{ESS} (Median) obtained with the Automatic Zig-Zag algorithm and \ac{HMC} algorithm given a pre-specified budget on bivariate Gaussian distributions and heavy-/ light-tailed distributions.}
	\label{t1}
	\begin{tabular}{r|cc}
		Target & Sampler&min Me ESS \\
		\hline
		\hline
		\texttt{IsoG2}& ZZ &1723\\
		& HMC & 2049 \\
		\hline
		\texttt{CorG2}& ZZ   & 317\\
		& HMC &  1419\\ 
		\hline
		\texttt{DscG2}& ZZ   &  261 \\
		& HMC &43\\ 	 \hline
		\texttt{BimodG2}& ZZ   & 185 \\
		& HMC &727\\ 	 \hline
		\texttt{LT2}& ZZ   & 1311\\
		& HMC &  2820\\ 	 \hline
		\texttt{HT2}& ZZ   &85\\
		& HMC &182 \\ 	
	\end{tabular}
\end{table}

The two algorithms performed very similarly on \texttt{IsoG2} (with a \ac{ESS} less then 20\% larger when \ac{HMC} was used), \ac{HMC} proved to be 4 to 5 times more efficient than Automatic Zig-Zag sampling on \texttt{CorG2}. Conversely, Zig-Zag sampling was 6 to 7 times more efficient than \ac{HMC} on \texttt{DscG2}. Despite the intrinsic advantage of \ac{HMC}, which is built to perform excellently on Gaussian targets, the observed comparable efficiency shows that the Automatic Zig-Zag sampling is competitive. 

With respect to robustness on these Gaussian targets, both algorithms performed well: the chains started in the mode reached the tails with adequate frequency and the chains initiated in the tails quickly converged towards the mode and continued to explore the target distribution. 

The performance was then tested against an heavy-tailed bivariate target (\texttt{HT2}) and a light-tailed bivariate target (\texttt{LT2}). The former is assumed to be distributed according to a bivariate Student-T with 2 degrees of freedom and the latter is assumed to have density $ p(\boldsymbol{x}) \propto e^{-\sum_{i=1}^{d}x_i^4/4 }$ for $ d= 2 $. \ac{HMC} was twice as efficient as Automatic Zig-Zag on \texttt{HT2}, whilst on \texttt{LT2}, \ac{HMC} was almost two times more efficient than Automatic Zig-Zag.

\begin{figure*}[h]
	\centering
	\begin{subfigure}[b]{0.48\textwidth}
		\centering
		\hspace*{-1cm}\includegraphics{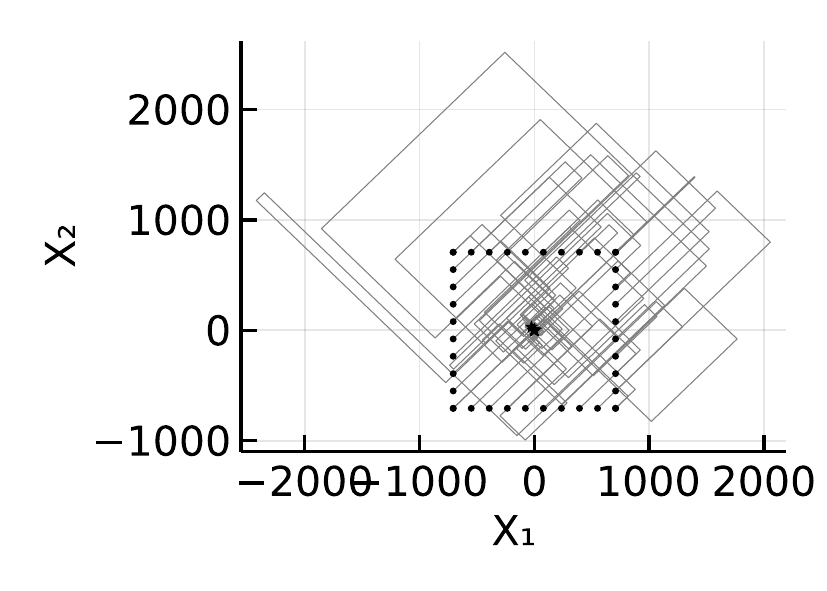}
		\caption{ Automatic Zig-Zag on \texttt{HT2}}
		\label{fig4.1a}
	\end{subfigure}	\begin{subfigure}[b]{0.48\textwidth}
		\centering
		\hspace*{-1cm}\includegraphics{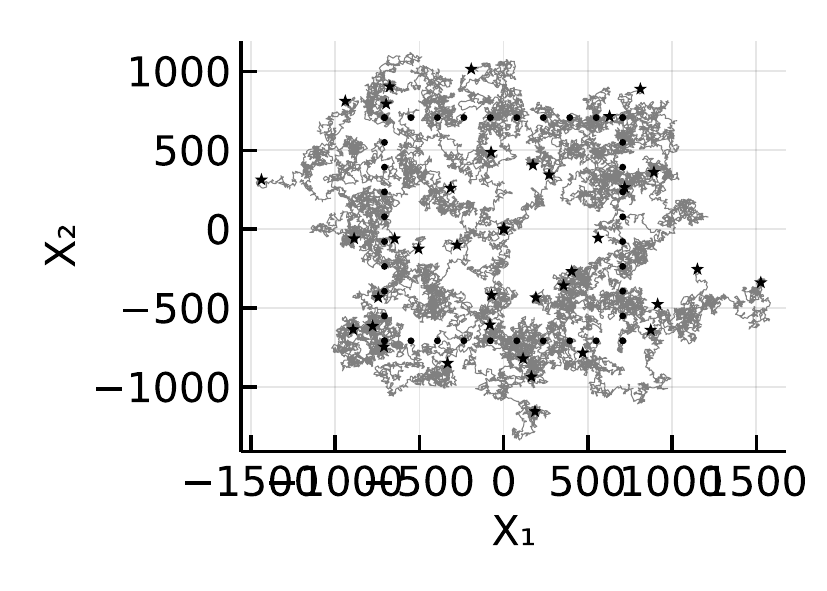}
		\caption{ HMC on \texttt{HT2}}
		\label{fig4.1b}
	\end{subfigure}\\
	\begin{subfigure}[b]{0.48\textwidth}
		\centering
		\hspace*{-1cm}\includegraphics{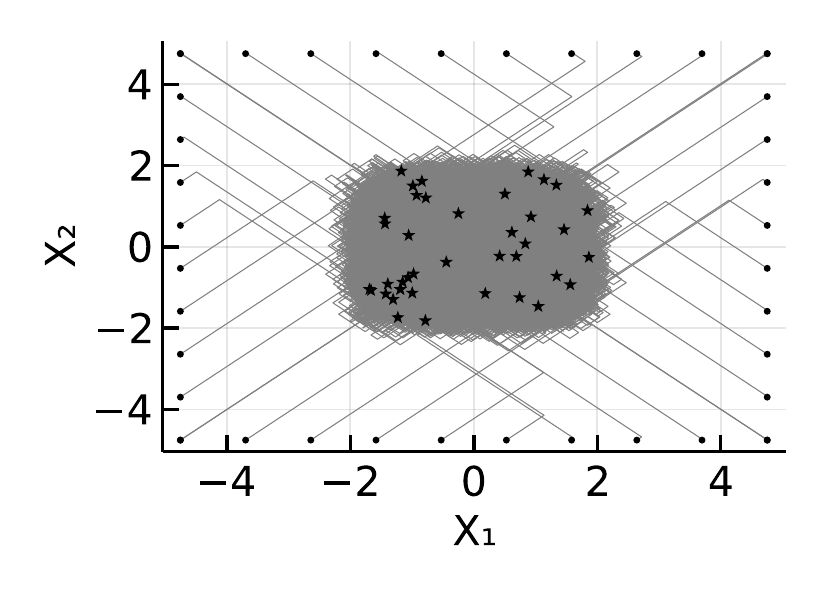}
		\caption{ Automatic Zig-Zag on \texttt{LT2}}
		\label{fig4.1c}
	\end{subfigure}
	\begin{subfigure}[b]{0.48\textwidth}
		\centering
		\hspace*{-1cm}\includegraphics{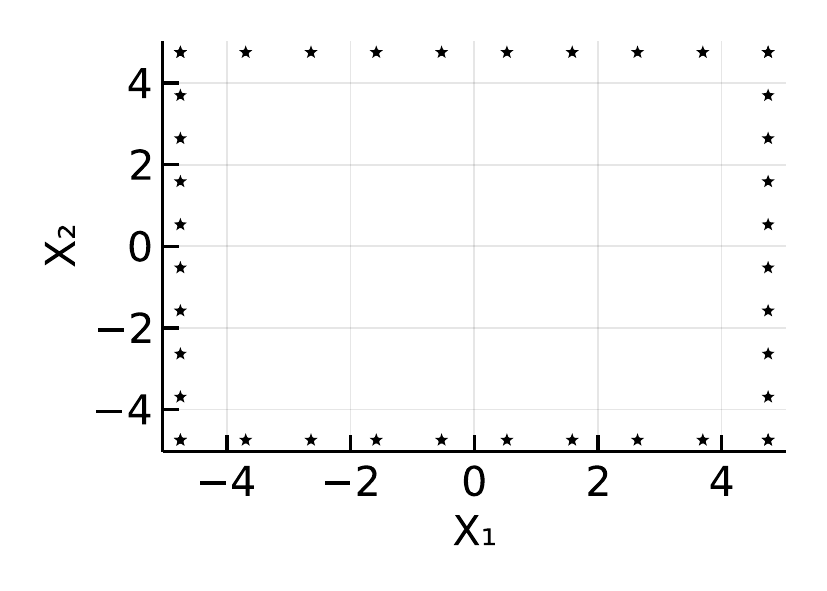}
		\caption{ HMC on \texttt{LT2}}
		\label{fig4.1d}
	\end{subfigure}
	\caption{Robustness of the algorithms on heavy- and light-tailed distributions: the two algorithms are tuned a priori and the chains are initiated at a grid of values in the tail of the distribution (dots). The Zig-Zag algorithm stops at 1000 skeleton points and HMC stops at 1000 iterations (final sample denoted by a star). When the chains don't move, such as in panel (d), the first and last sample overlay. }
	\label{fig4.1}
\end{figure*}

The Automatic Zig-Zag algorithm however, proved to be more robust to these two examples providing consistent exploration of the tails in \texttt{HT2} and fast convergence towards the mode when starting in the tails for both \texttt{HT2} and \texttt{LT2}. These are reported graphically in Figure \ref{fig4.1} where multiple chains starting from a grid of values in the tails of the distribution were run for a limited number of iterations/skeleton points. In Figure \ref{fig4.1a} and \ref{fig4.1c} the rapid convergence towards the mode of the Zig-Zag algorithm can be appreciated. Conversely, the \ac{HMC} chains struggled to move towards the mode of the heavy tailed distribution (Figure \ref{fig4.1b}) and did not move at all on the light-tailed distribution (Figure \ref{fig4.1d}): the gradient in these locations suggested proposals far off in the opposite tail which were then never accepted. 

Comprehensive results from the simulation study, including illustrations of the optimality of the tuning of the  Zig-Zag algorithm, are reported in Section S3 of the Supplementary Information. 

\section{Real data applications }\label{s.5}
In this section, some examples of the application af Automatic Zig-Zag sampling to real data analyses are proposed. The first is an example of a non-linear regression model from a Bayesian Methods textbook \citep{carlin2008bayesian}; and the second example is a parametric survival model.
\subsection{A textbook example}
We reproduce the analysis of \cite[page 176]{carlin2008bayesian}, which analyses data on dugongs (sea cows), considering a non-linear growth model to relate their length in meters ($ Y_j$) to their age in years ($ z_j $). 
The model assumed is:
\begin{equation}\label{key}
	Y_j=\alpha-\beta\gamma^{z_j}+\varepsilon_j \qquad \text{for } j=1, \dots, J
\end{equation}
with normally distributed errors $ \varepsilon_j\overset{iid}{\sim }N(0, \sigma^2)$. 

The parameters are $ \alpha>0, \beta>0, 0\leq\gamma\leq1, \sigma>0 $; the parameters are explored on the following transformed space:
\begin{equation}\label{key}
	\begin{split}
		x_1&= \log(\alpha)\\
		x_2&= \log(\beta)\\
		x_3&= \log\left( \frac{\gamma}{1-\gamma}\right)\\
		x_4&= \log(\sigma).
	\end{split}
\end{equation}
The priors are assumed flat on their original domain except for $ \gamma $ which has a Beta$(7,7/3)$ prior. This model presents some challenges in that this parametrization favours correlation in the posterior distribution and different scales for the parameters. 

\begin{figure}[h]
	\centering
	\hspace*{-1cm}\includegraphics{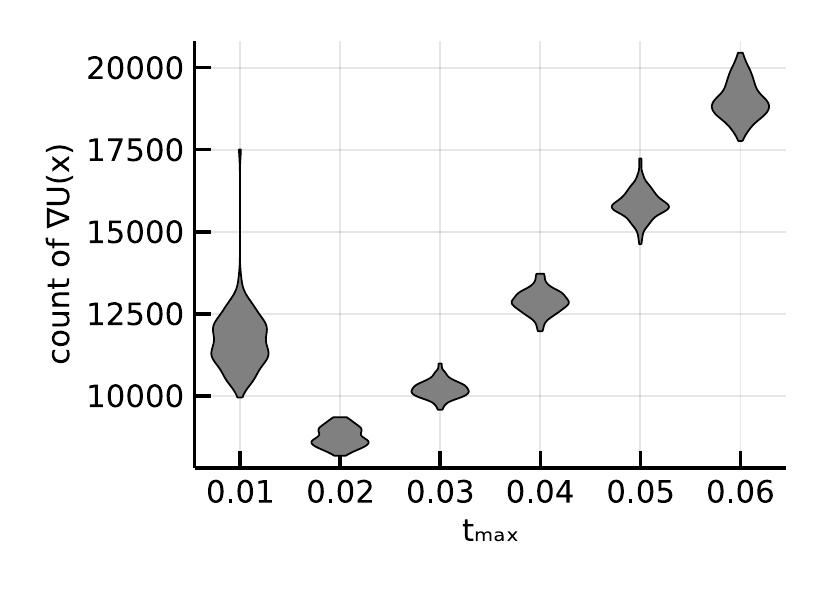}
	\caption{Selection of $ t_\textsc{max} $ via preliminary runs. Violin plots of the total number of gradient evaluations form 100 samples of 1000 skeleton points at given values of $ t_\textsc{max} $. }
	\label{fig5.2}
\end{figure}

The selection of an appropriate $ t_\textsc{max} $ was done via preliminary runs, as described in Section \ref{s.3}, that guided the choice of an efficient value at $ t_\textsc{max}=0.02 $ (see Figure \ref{fig5.2}).

\begin{figure}
	\begin{subfigure}[b]{0.48\textwidth}
		\centering
		\hspace*{-.5cm}\includegraphics[scale=0.5]{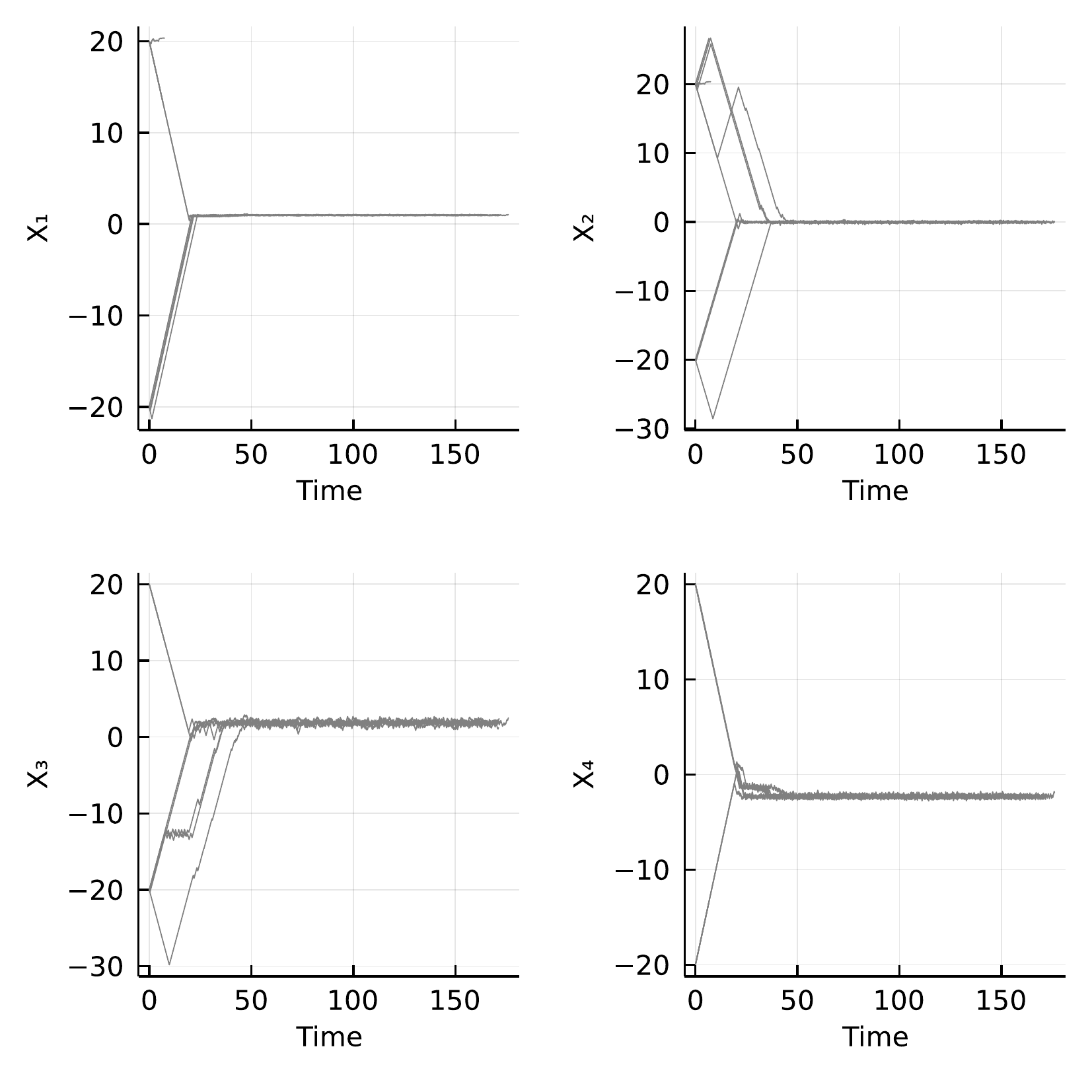}
		\caption{Automatic Zig-Zag skeleton}
		\label{fig5.5a}
	\end{subfigure}
	\begin{subfigure}[b]{0.48\textwidth}
		\centering
		\hspace*{-.5cm}\includegraphics[scale=0.5]{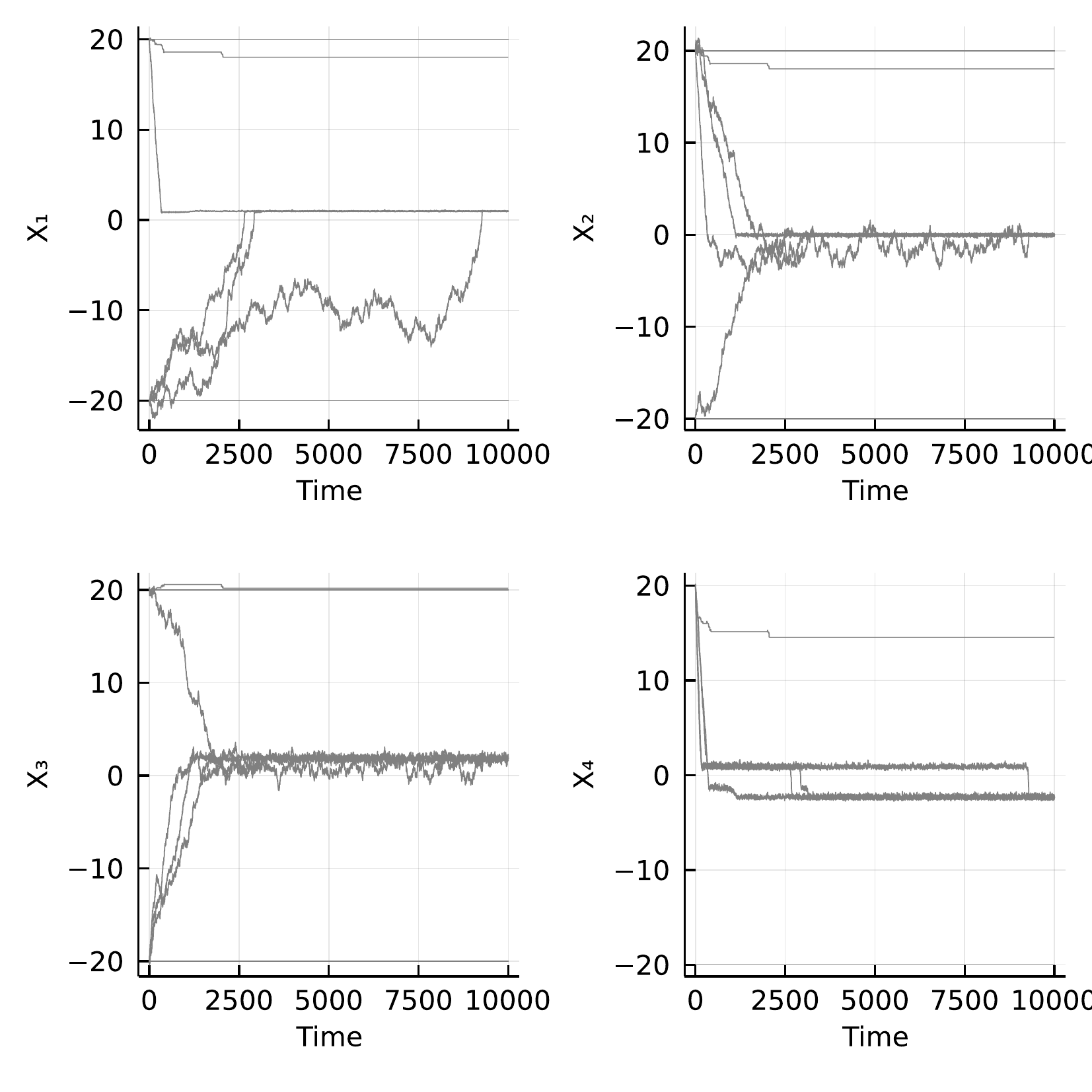}
		\caption{HMC samples}
		\label{fig5.5b}
	\end{subfigure}
	\caption{Skeleton plot (Fig. \ref{fig5.5a}) and \ac{HMC} chains (Fig. \ref{fig5.5b}) for the 4 parameters when initiated in location not close to the mode: while Zig-Zag rapidly converges to the true value, \ac{HMC} takes much longer to reach convergence or, in some case does not move.}
	\label{fig5.5}
\end{figure}

The comparative results against \ac{HMC} showed the same pattern observed in Section \ref{s.4}: \ac{HMC} was slightly faster than Zig-Zag in exploring the space, leading to an increased \ac{ESS} given a limited budget
. Zig-Zag however was much more robust to the choice of initial values: it was able to reach the mass of the distribution very quickly. Conversely \ac{HMC} often remained stuck in initial values (or in other values away from the mode), struggling to reach convergence (Figure \ref{fig5.5}). This behaviour was also observed when more elaborate adaptations of \ac{HMC} were used, such as the Non U-Turn Sampler \citep{hoffman2014no} implemented in the software \texttt{Stan} \citep{carpenter2017stan}.

\subsection{Parametric survival model}
Automatic Zig-Zag was tested on the inference of a  Bayesian parametric survival regression model fitted to a sample of individuals from a large synthetic database \citep{simulacrum}. The whole data are described below and a model was fitted initially to a sample of 500 individuals. The dataset is analysed in full in the next section, where automatic super-efficiency is explored.

\subsubsection{Data}
The dataset comprises information on 2,200,626 synthetic patients and their 2,371,281 synthetic tumours, including the time of each cancer diagnosis, the time/type of final event observed (i.e. time of death if dead or censoring time if alive), basic demographics of the patients and on their tumour history (e.g. time of surgery if surgically addressed, therapy type and timings). 

A parametric survival regression model \citep{jackson2016flexsurv} was fitted to these data in order to explain the survival-time from first tumour diagnosis with few individual-specific covariates. Note that the results reported here should not be interpreted as real, not only because the data used are synthetic, but also because the effects estimated here should be corrected for other covariates which were not included in this analysis and are known to affect and confound survival from diagnosis. Other simplifying assumption were made, including uninformative missingness, uninformative loss to follow-up and no left censoring. Thanks to the high  completeness of the dataset only 2,565 patients were excluded due to missing at least one key variable (i.e. date/type of final outcome). 

The final dataset analysed consisted of: a set of times $ t_j $ from diagnosis of the first tumour to either death or censoring; a set of event type $ c_j $, with $ c_j=1 $ for death and $ c_i=0 $ for (administrative) censoring; and a set of covariates $ z_j^1, \dots, z_j^g $ for $ j=1,2,\dots, J $, with $ J $= 2,198,061 individuals. 

\subsubsection{Model}
A Weibull model was assumed, i.e. the time to death has probability density function:
\begin{equation}\label{e17}
	f(t; \mu, \alpha) = \frac{\alpha}{\mu}\left(\frac{t}{\mu}\right)^{\alpha-1} e^{-\left(\frac{t}{\mu}\right)^\alpha}
\end{equation}
and survival function:
\begin{equation}\label{key}
	S(t; \mu, \alpha)=	e^{-\left(\frac{t}{\mu}\right)^\alpha}
\end{equation}
so that the overall likelihood of the vectors of outcomes $ \boldsymbol{t}= t_1, t_2, \dots, t_J$ and $ \boldsymbol{c}=c_1, c_2, \dots, c_J$ is:
\begin{equation}\label{key}
	\ell(\boldsymbol{t}, \boldsymbol{c}\vert\mu, \alpha)=\prod_{j=1}^J	\mathbb{I}_{c_j=1}f(t; \mu, \alpha) + \mathbb{I}_{c_j=0}	S(t; \mu, \alpha) 
\end{equation}
The scale parameter $ \mu $ was related to the covariate of interest $ z^1, \dots, z^g $ via log link:
\begin{equation}\label{e20}
	\log(\mu_j)= \beta_0+\beta_1z_j^1+\dots +\beta_gz_j^g
\end{equation}
Let $ z^1_j $ be the age at diagnosis of patient $ j $, and $ z^2_j $ be the discrete variable identifying the spreading status of the cancer: if $ z^2_j =0$, the cancer of patient $ j $ haven't spread to other sites (i.e. it is in stage 2 or smaller)  if $ z^2_j =1$, the cancer of patient $ j $ is likely to have spread to other sites (i.e. it is in stage 3 or greater).

In the  Zig-Zag notation, the location vector $	\boldsymbol{ X }  $ was then composed by all the parameters of the model:
\begin{equation}\label{key}
	\boldsymbol{ X }=\left(\log(\alpha), \beta_0, \beta_1, \beta_2 \right).
\end{equation}

\subsubsection{Results for 500 individuals}
A randomly selected subset of $ J=500 $ individuals was initially analysed. 

In this model, the parameter space is slightly unbalanced: the first component ($ \log(\alpha) $) highly affects the shape of the potential, constraining all the other components, hence the \ac{MCMC} is doomed to mix slowly overall. This ill-behaviour is a combination of two aspects explored in the simulations of Section \ref{s.4}: the components of $ \boldsymbol{X} $ have different scales and are highly correlated. 

\begin{figure}
	\centering
	\hspace*{-0.5cm}\includegraphics{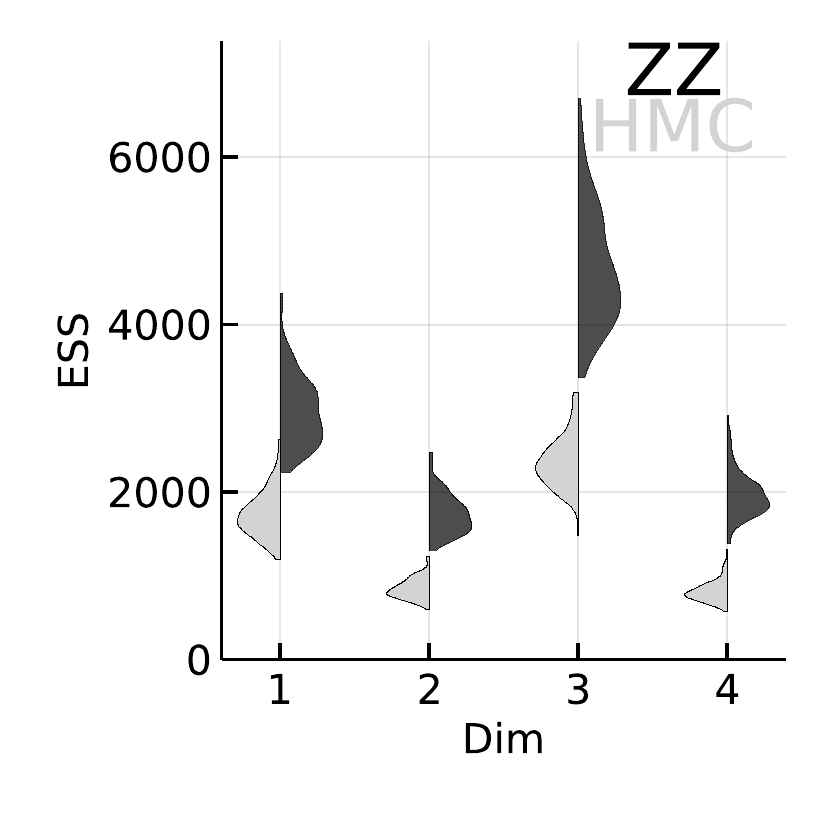}
	\caption{Violin plot of the \ac{ESS} of each dimension obtained on 100 simulated chains of \ac{HMC} and Automatic Zig-Zag given a pre-sepcified budget of 200000 gradient evaluations. }
	\label{fig5.6}
\end{figure}

The Zig-Zag sampler performed satisfactory in exploring this challenging target distribution: it was shown to be more robust than a properly-tuned \ac{HMC} (results reported in Section S4 of the Supplementary Information). Moreover, the Zig-Zag sampler was shown to be more efficient than \ac{HMC}, achieving systematically higher \ac{ESS} on all dimensions as reported in Figure \ref{fig5.6}.

These results come from the analysis of a small subset of the population but, as more data are included, the evaluation of the likelihood and its gradient becomes more and more expensive, and the overall exploration of the space is slower. This motivates the need to exploit super-efficiency which is described in Section \ref{s.6} in a general context. Results from the analysis of the full dataset using our super-efficient Zig-Zag sampler are presented in Section \ref{s6.4}.

\section{Automatic super-efficiency}\label{s.6}
One of the most appealing properties of the Zig-Zag algorithm, and of \acp{PDMP} more generally, is super-efficiency. An algorithm is defined to be \textit{super-efficient} if it ``is able to generate independent samples from the target distribution at a higher efficiency than if we would draw independently from the target distribution at the cost of evaluating all data" \citep{bierkens2019zig}.

\subsection{Subsampling}
Super-efficiency can be obtained if the potential $ U(\boldsymbol{x}) $ takes a particular form.
Specifically, consider $ U(\boldsymbol{x}) $ for which $ \partial_i U(\boldsymbol{x})=\frac{\partial U(\boldsymbol{x})}{\partial x_i} $ admits representation:
\begin{equation}\label{ess1}
	\partial_i U(\boldsymbol{x})=\frac{1}{J}\sum_{j=1}^{J} E_i^j (\boldsymbol{x}) 
\end{equation}
for $ i=1, \dots, d $. 
This representation is available, for example, when the target density can be factorised in a series of $ J $ components (e.g. a sum of $ J $ observation-specific likelihoods of \ac{iid} observations). 

With representation \eqref{ess1}, the following steps allow the construction of an algorithm to sample from the correct target distribution.
\begin{enumerate}\setlength{\itemsep}{0mm}
	\item Define a dimension-specific collection of switching rates (with $ i=1, \dots, d $ indexing the dimension), where each element of the collection can be thought of as the observation-specific factor of the potential (with $ j=1, \dots, J $ indexing the observation):
	\begin{equation}\label{e22}
		m^j_i(t):= \max\left\lbrace v_i E_i^j (\boldsymbol{x}(t)), 0  \right\rbrace    
	\end{equation}
	for $  i=1, \dots, d $; $ j=1, \dots, J $.
	\item Find a collection-specific function $ M_i(t)  $ which bounds all the rate of a specific dimension $ i $:
	\begin{equation*}
		m^j_i(t)\leq M_i(t)  \qquad \text{for all } j=1, \dots, J 
	\end{equation*}
	for $  i=1, \dots, d $. This bound can vary over time $ t $ or be constant, i.e. $ M_i(t)=c_i $.
	\item Sample the first event time from $ d $ homogeneous Poisson processes: $ \tau_i \sim PP( M_i(t) ) $ and take:
	\begin{equation*}\begin{split}
			\tau &= \min\left\lbrace \tau_1, \tau_2, \dots, \tau_d \right\rbrace \\
			i_0 &= \text{argmin}\left\lbrace \tau_1, \tau_2, \dots, \tau_d \right\rbrace .
		\end{split}
	\end{equation*}
	\item Sample an index of the observations: $$ j_0\sim \text{Uniform} (1, 2, \dots, J) .$$
	\item Accept the switch for dimension $ i_0 $ with probability $ m^{j_0}_{i_0}(\tau)/ M_{i_0}(\tau) $. 
\end{enumerate}
The process of using only one observation (or, any other unbiased estimator of $ 	\partial_i U(\boldsymbol{x}) $ in \eqref{ess1} which uses less than $ J $ computations) is called \textit{subsampling}. 
Subsampling as described above (i.e. when only one observation is used) allows to reduce computational complexity of the algorithm by a factor $ O(J) $. This result has been proven in \cite{bierkens2019zig} and a few considerations were drawn: the resulting chain mixes more slowly than a chain obtained with the non-subsampling algorithm; nevertheless, control variates can be used to further improve the efficiency of the Zig-Zag with subsampling. 

A straightforward way to extend the methods presented in Section \ref{s.3} is to allow the input to be directly the observation-specific density $ E^i_j $, with the formulation of a generic potential which depends on the observation index $ j $.

\subsubsection{Challenges}

To properly implement subsampling, a collection-specific upper bound $ M_i(t) $ (or a constant bound $ c_i $) must be available, but in a generic example it may not be possible to find a bound analytically. With the introduction of an automatic method, all the functional information on the derivatives of the potential is lost. 

To address this issue, a constant-local approach is again adopted: it would be sufficient to find a value $ c_i $ for given starting values $ (\boldsymbol{v_s}, \boldsymbol{x_s}) $ within an horizon of length $ t_\textsc{max} $: $ (t \in [0,t_\textsc{max}]) $, so for a specific dimension $ i $ the bounding rate would be:
\begin{equation}
	m_i^j(t) \leq c_i \qquad \forall j=1, 2, \dots J  \text{ for } t\in (0, t_\textsc{max})	.
\end{equation}
{If this approach is taken, $ c_i $ refers specifically to the starting values $ (\boldsymbol{v_s}, \boldsymbol{x_s}) $ and a new $ c_i $ should be considered whenever a switch or a deterministic move is made. }Even within this horizon {$ [0,t_\textsc{max}] $}, however, finding a maximum by evaluating and maximizing all the $ J $ observation-specific rates and then comparing them would be counter-productive: all the gain of super-efficiency would be lost in this optimization step. A super-efficient method to overcome this challenge is proposed below.

\subsection{Bounding unknown rates}\label{s6.2}
The main idea of our proposal to find an efficient estimate $ \widehat{c}_i  $ of $ c_i $ is to consider only a small sample of size $ q  $ of the available switching rates, maximise them to obtain a sample of rate-specific maxima/bounds and finally apply extreme-value theory methods to infer the population maximum across all the rates. 

Given a local starting point $ (\boldsymbol{v_s}, \boldsymbol{x_s}) $ and within a given horizon of length $ t_\textsc{max} $, an estimate $ \widehat{c}_i $ of $ c_i $ is obtained with the following steps: 
\begin{enumerate}
	\item select a sample $ \mathcal{Q} $ of size $ q $ from the $ J $ rates available in the collection;
	\item run a numerical optimization algorithm (e.g. our version of Brent's method) to obtain rate-specific maxima of the $ q\times d $ dimension-specific sampled rates: $$\overline{\lambda_i}^j=\max_{t \in(0, t_\textsc{max})} m_i^{j}(t)  ;$$ for $ j\in \mathcal{Q} $, for $ i=1,\dots d $;
	\item for each dimension $ i=1,\dots,d $, use the $ q $ values of $\overline{ \lambda_i}^j $ to fit a \ac{GPD} and  obtain estimated parameters $ \hat{\xi}_i, \hat{\sigma}_i $ of the \ac{GPD};
	\item use the parameters to predict $ c_i $ in a \textit{return value} perspective: $\widehat{c}_i= q^{(i)}_{\frac{J-1}{j}} $, with $ q^{(i)}_{\frac{J-1}{j}} $ the $ 1-1/J $th quantile of the extreme value distribution with parameters $ \hat{\xi}_i, \hat{\sigma}_i $; for each dimension $ i=1,\dots, d $. 
\end{enumerate}
The estimated $ \widehat{c}_i $ can be then considered as the \textit{population bound}: the value that would be the maximum (the only one at or above its value) if we had a sample of size $ J $.

More detail on the results used from extreme-values is reported in the Supplementary Information, Section S5. {If the acceptance step of the subsampling algorithm shows that $ \widehat{c}_i $ is found not to bound some rates, than a new set of rates $ \mathcal{Q} $ is drawn and steps 2 to 4  above are run again. }

\subsection{Practical considerations}
The method proposed in Section \ref{s6.2} still retains the \textit{automatic} flavour of the algorithms proposed here but allows to exploit what is thought of as the most-promising property of Zig-Zag samplers and other \acp{PDMP}. In implementing this idea in practice, however, a few choices must be made. 

Firstly, one should decide on the \textit{level of super efficiency} desired: one of many \ac{iid} observations already provides a unbiased estimate for the rate, but it might be better to include more{, say $ h $,} observations in order to have a more-representative sample of the population. {The larger $ h $, the more homogenous the subsample-specific rates are. As a consequence the process mixes better as the subsample-specific rates resemble better the population rate. When one, or very few observations contribute to each subsample-specific rate, the process will switch often reflecting the heterogeneity across them. } 

Likewise, $ q $, i.e. the number of rates that are selected for the estimation of the bounds, highly affects the quality of the estimator $ \widehat{c}_i $, which, if underestimated, could lead to the samples from the ZZ sampler being overdispersed with respect to the target distribution. A robustness factor $ r\geq1 $ is introduced so that the upper bound is effectively larger than the predicted return value by the Generalised Pareto: $\widehat{c}_i=r\times q^{(i)}_{\frac{J-1}{j}} $.

These quantities: the number of observations per rate, $ q $, and $ r $, should be considered tuning parameters and chosen on a case-by-case basis via preliminary analysis as exemplified in the following section. {For example, finding that that the rates exceed their estimated bound $ \widehat{c} $ often, suggests that $ r $ might have to be increased. }

Lastly, note that, while $ q $ rates are needed to infer $ \widehat{c} $, the optimization routine on each of these rates could be parallelised: the $ q  $ maxima $ \overline{\lambda_i}^j $ can be computed independently, enabling even higher efficiency.
\subsection{Parametric survival model on big data}\label{s6.4}
In this section we fit the Parametric survival model of Equations \eqref{e17}-\eqref{e20} to the total population of $ J=2,198,061 $ individuals. 

As a staring point, we attempted the more computationally-expensive approach of using the standard Automatic Zig-Zag algorithm, whose results are reported in blue in Figure \ref{fig6.1}. To obtain such a skeleton (composed by 5000 switching times), circa 63,000 gradient evaluations were made, each of which is a computation of order $ J\approx2  $ million. 
The overall clock time elapsed was 4 hours, after careful tuning of $ t_{\textsc{max}}$. 
\subsubsection{Subsampling setup}
We ran the Automatic Zig-Zag with subsampling, where at each iteration the rate was approximated by considering a potential that accounts for fewer than $ J $ observations. 

When we implemented the most drastic subsampling, using only one observation and  we approximated  the rate by $ m_i^j(t) $ as defined in \eqref{e22}, many problem arose. The switching rates were very heterogeneous in the observation that was (sub-)sampled which meant that it was difficult to estimate appropriate bounding constants $ \widehat{c}_i $ that were valid for the whole population; for the same reason mixing was very slowly. 

We therefore decided to use subsamples of size $ h>1 $ to approximate the rates. Let $ \mathcal{S}_l $ be a sample of size $ h $ of indexes, drawn without replacement from $ \{1, 2, \dots, J\} $. The rates are generated using estimates for the potential of the type: 
\begin{equation*}
	E^l_i(\boldsymbol{x}) = \sum_{j\in\mathcal{S}_l} \frac{E_i^j}{h}.
\end{equation*}

A subsample size of $ h=20 $ gave very satisfactory results with robust estimates of $ \widehat{c} $ and good mixing.

The estimates of $ \widehat{c}  $ were obtained by evaluating $ q=1000 $ other rates, whose maxima were used to fit a \ac{GPD}; the robustness factor was set to $ r=2 $.
\subsubsection{Results}
\begin{figure}
	\includegraphics[scale=0.22]{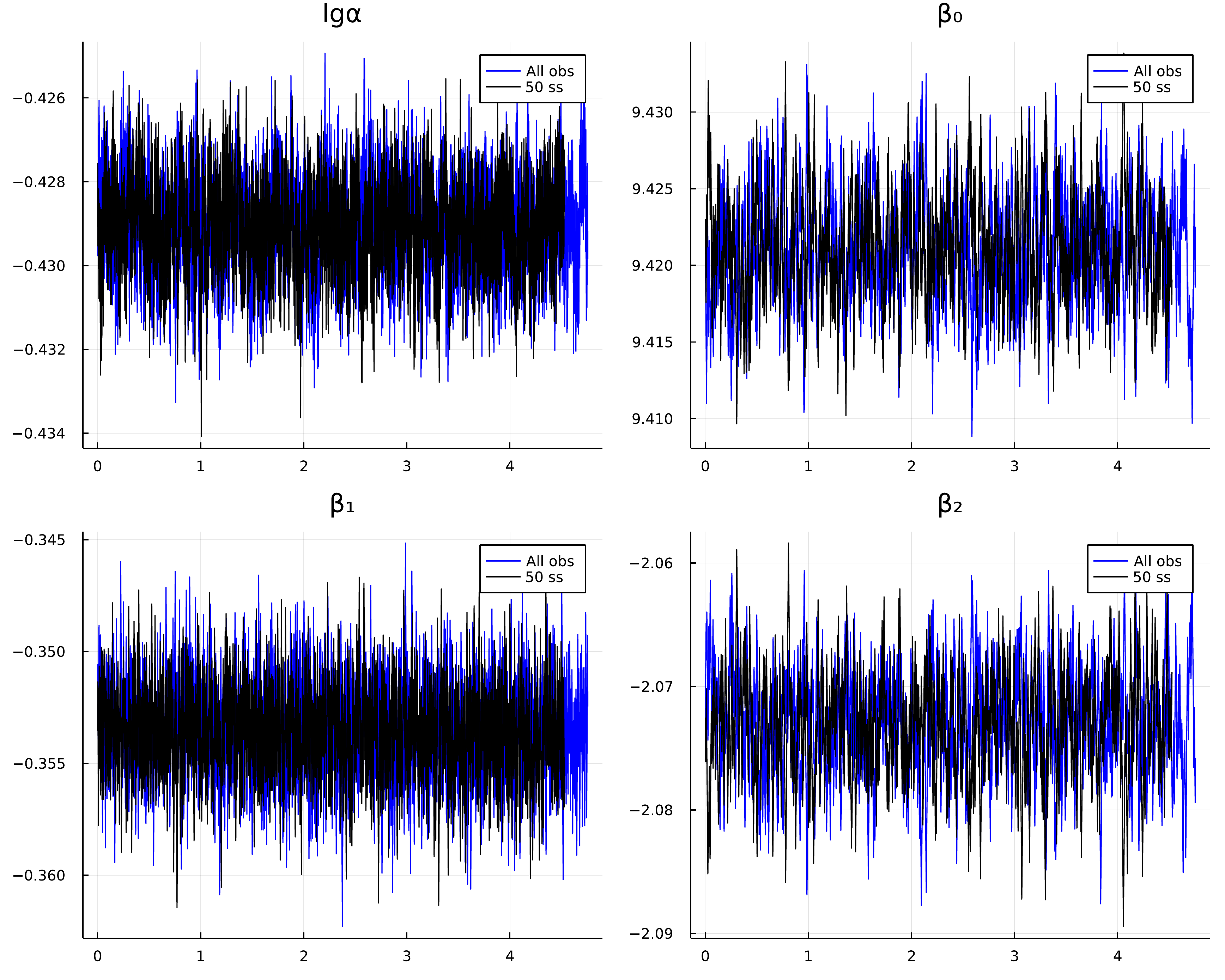}
	\caption{Skeleton of the Zig-Zagprocess without subsampling (black) and with subsampling (blue).}
	\label{fig6.1}
\end{figure}

The resulting 5000 switching-point skeleton appears to have mixed well and converged to the same distribution as its expensive, non-subsampling counterpart (see Figure \ref{fig6.1}).

The overall clock-time elapsed is circa 30 minutes, hence the gain from applying the subsampling techniques is tangible: our implementation of the subsampling technique was 7 to 8 times faster than the standard method. The algorithm was run without  any parallelaziation in the estimation of the local upper bounds, hence an even shorter computation time could be achieved. Moreover, because the implementation of the Automatic Zig-Zag with subsampling runs substantially faster, a more precise estimate of the optimal $ t_\textsc{max} $ can be produced from the same computational budget. For the full data it was almost impossible to accurately tune $ t_\textsc{max} $, given the long computation time, and our initial guess led to a run-time of 24 hours, which was reduced to 4 hours only after using the $ t_\textsc{max} $ obtained from the pilot runs of the sub-sampling algorithm.

While the choice of the level of subsampling $ h $ was done by trial and error, it is a straightforward process that, thanks to the speed of the algorithm, can be performed a priori.  Notably, even if the Automatic Zig-Zag with subsampling requires a more accurate tuning of the parameters $ h, r$ and $q $, it still retains the automatic properties that the original algorithm has, since no further information on the shape or properties of the target distribution were used. 

\section{Discussion}\label{s.7}
The theory behind \acp{PDMP} is developing quickly and forming a substantive body of results that make \ac{PDMP}-based algorithms extremely promising. 
Little work exists on the use of these algorithms to address applied problems, with notable exceptions including: variable selection problems \citep{chevallier2020reversible}, inference of diffusion bridges \citep{bierkens2021piecewise}, and inference of phylogenetic trees \citep{koskela2022zig}. 
These applications develop bespoke versions of the Zig-Zag sampler, and other \ac{PDMP}-based algorithms, and demonstrate their usefulness and efficiency within the specific applications considered. 

Generalisations of \ac{PDMP} algorithms that make them applicable in any context are even more rare: the simulation of a \ac{PDMP} is strictly constrained by the availability of adequate upper bounds of the switching rate or by closed-form solutions to the integral of the rate for the time-scale transformation.  To our knowledge, there are only two papers that provide a general tool to draw samples using \acp{PDMP} requiring only the evaluation of the gradient of the target density. The \ac{NuZZ} \citep{pagani2020nuzz} uses numerical integration to simulate the next switching event by time-scale transformation. {The numeric integrator requires the evaluation of the rate $ \lambda(t) $ for a grid of values for $ t $ (from 7 to 14 points), and it is computed at each iteration of a root-finding method that derives the switching time ($ \tau $ in Equation \ref{eq2.8}). While there might be cases when the \ac{NuZZ} is the most efficient solution, we have found that its numerical routine requires more evaluations of $ \lambda(t) $ per switching point compared to our algorithm, whose optimization method resulted extremely efficient, requiring often only 4 evaluations of $ \lambda(t) $; the appropriate tuning of $ t_{\textsc{max}} $ keeps the total number of Poisson Process proposals for thinning small and, in the best cases, around 1. Lastly, the \ac{NuZZ} is, differently from ours, an approximated algorithm, whose error diminishes as the number of points used for the numerical integrator increases. }

Another simulation scheme for \acp{PDMP} is proposed in \cite{bertazzi2021approximations}, which solves the same problem by exploiting Euler approximations of the switching rate, abandoning {once again} exactness for the sake of generalizability. {Similarly to the NuZZ, approximation schemes require the evaluation of $ \lambda(t) $ for a grid of values, jeopardizing efficiency. }

Our work instead welcomes an intensive use of modern \ac{AD} techniques, which allow the exploration of any target whose (minus log) density is differentiable. Rates computed via \ac{AD} are matched with a numeric optimization method that allows the quick computation of a local upper bound to sample the switching time via thinning. The resulting Automatic Zig-Zag sampler  provides a robust and general way to sample from any distribution with differentiable {log-}density without the need of any further information. We tested Automatic Zig-Zag and showed it to be competitive with \ac{HMC}: although \ac{HMC} is often apparently more efficient, we found it to be considerably less robust when more challenging situations are presented and {when} starting values are far from the support of the target distribution.
On most of the real-data scenarios presented, the Automatic Zig-Zag sampler was shown to be superior to \ac{HMC} providing robust results with a simpler tuning process. 

In addition to automation of the differentiation and upper bound calculation, the Automatic Zig-Zag sampler has been further extended to benefit from super-efficiency, the most appealing property of \ac{PDMP} samplers. The power of super-efficiency in this context has been demonstrated in practice on the analysis of a large dataset. 

Automatic Zig-Zag presents only a few limitations, the first of which is the use of a numeric method to determine a local upper-bound on the switching rate. 
As most of the available optimization methods, Brent's optimization (and our modified method) does not guarantee convergence to a global maximum in the interval considered. Nevertheless, we have found that in practice the method is robust and it rarely fails on the type of functions that need to be bounded in the Zig-Zag algorithms and, given its low computational burden, we were able to introduce further checks to prevent avoidable errors in the computation of the upper bound. Moreover, the tuning parameter $ t_\textsc{max} $, i.e. the width of the interval over which the optimization is run, can be reduced to decrease the probability of optimization failure. {A similar consideration applies to the method presented in Section 6 which lacks guarantees that the estimator $ \widehat{c} $ would bound all the rates. Nevertheless, we again introduce checks and parameters that can make automatic super-efficiency more robust. }Another limitation of the work presented here is that it contemplates only smooth densities on unbounded domain. The general question of the behaviour if \acp{PDMP} on piecewise-smooth and bounded densities is addressed in \cite{chevallier2021pdmp}, however, the results presented were derived using the knowledge of the discontinuities in smoothness and on the bound, hence they are not applicable in a general context. 

Another possible improvement to the Automatic Zig-Zag sampler is the adaptation of the velocity space to the target density considered, similarly to \cite{bertazzi2020adaptive}. This would improve the general performance of the algorithm, not only in the aspects described by \cite{bertazzi2020adaptive}, but also it should lead to a choice of $ t_\textsc{max} $ that is homogeneously optimal for all dimensions. Progress in this direction is the focus of our future work. 

Lastly, while a supplementary code of this paper is provided and contains useful functions to understand and replicate our methods, a full package that implements the Automatic Zig-Zagsampler for Bayesian analyses is being developed to make this method usable by practitioners in all settings. 

The availability of a continuous-time algorithm that provides samples from a desired  target requiring only a functional form for its (minus log) density opens several possibilities for probabilistic programming languages, substantially advancing the current state of the art.
In this paper we have made contributions which facilitate the use of \ac{PDMP} methods on a substantially expanded family of targets, and we hope that our work can therefore greatly expand the wide applicability of \acp{PDMP}.

\section*{Supplementary information}
Supplementary Information is available in the form of a report attached to the paper.
\section*{Acknowledgments}
This work has been funded by EPSRC grant EP/R018561/1, New Approaches to Bayesian Data Science: Tackling Challenges from the Health Sciences. {GOR is further supported by The Alan Turing Institute. }

The authors are grateful to professor Jonathan Tawn for his insights and comments on  the extreme-values methods used in Section \ref{s.6}. 

\section*{Disclosures and Declarations}

Data and model for the case study in Section \ref{s.5} is available in \cite{carlin2008bayesian}, (page 43). 

The case-study in Sections \ref{s.6} and \ref{s.5}  uses artificial data from the Simulacrum, a synthetic dataset developed by Health Data Insight CiC derived from anonymous cancer data provided by the National Cancer Registration and Analysis Service, which is part of Public Health England \cite{simulacrum} and is available to download from \href{https://simulacrum.healthdatainsight.org.uk/using-the-simulacrum/requesting-data/}{https://simulacrum.healthdatainsight.org.uk/using-the-simulacrum/requesting-data/}. Data are artificial hence no Ethical approval or Consent to participate is needed.

All code used to generate results and plots is available online at 
\href{https://github.com/alicecorbella/ZZpaper}{https://github.com/alicecorbella/ZZpaper}.

\printbibliography

\begin{acronym}
\acro{MCMC}{Markov chain Monte Carlo}
\acro{HMC}{Hamiltonian Monte Carlo}
\acro{BPS}{Bouncy Particle Sampler}
\acro{PDMP}{piecewise deterministic Markov process}
\acro{MALA}{Metropolis-adjusted Langevin algorithm}
\acro{AD}{Automatic Differentiation}
\acro{rv}{random variable}
\acro{NHPP}{non-homogeneous Poisson process}
\acro{HPP}{homogeneous Poisson process}
\acro{ESS}{Effective Sample Size}
\acro{ACF}{autocorellation function}
\acro{GEV}{Generalised Extreme Value}
\acro{GPD}{Generalised Pareto Distribution}
\acro{POT}{peak over threshold}
\acro{NuZZ}{Numeric Zig-Zag}
\acro{GPD}{Generalised Pareto Distribution}
\acro{iid}{independent and identically distributed}
\acro{pdf}{probability density function}
\end{acronym}
\end{document}